\documentclass[aps,showpacs,amsfonts,amsmath, amssymb,twocolumn,floatfix]{revtex4}
\usepackage[final]{graphicx}
\usepackage{xcolor}
\usepackage[colorlinks=true, linkcolor=blue, citecolor=blue, urlcolor=blue]{hyperref}
\usepackage{verbatim}
\usepackage{csquotes}
\usepackage{braket}
\usepackage{mathtools}

\newcommand{\vct}[1]{\mathbf{#1}}

\renewcommand\Re{\operatorname{Re}}
\renewcommand\Im{\operatorname{Im}}
\newcommand\Tr{{\rm Tr}}

\newcommand{\be}{\begin{equation}}
\newcommand{\ee}{\end{equation}}
\allowdisplaybreaks

\DeclareSymbolFont{bbgreek}{U}{bbold}{m}{n}
\DeclareMathSymbol{\bbmu}{\mathbb}{bbgreek}{'26}
\DeclareMathSymbol{\bbeps}{\mathbb}{bbgreek}{'17}

\begin{document}

\title{Heat radiation and transfer for point particles in arbitrary geometries}

\author{Kiryl Asheichyk}
\email[]{asheichyk@is.mpg.de}
\author{Boris M\"uller}
\author{Matthias Kr\"uger}

\affiliation{4th Institute for Theoretical Physics, Universit\"at Stuttgart, Pfaffenwaldring 57, 70569 Stuttgart, Germany}
\affiliation{Max Planck Institute for Intelligent Systems, Heisenbergstrasse 3, 70569 Stuttgart, Germany}

\begin{abstract} 
We study heat radiation and heat transfer for pointlike particles in a system of other objects. Starting from exact many-body expressions found from scattering theory and fluctuational electrodynamics, we find that transfer and radiation for point particles are given in terms of the Green's function of the system in the absence of the point particles. These general expressions contain no approximation for the surrounding objects. As an application, we compute the heat transfer between two point particles in the presence of a sphere of arbitrary size and show that the transfer is enhanced by several 
orders of magnitude through the presence of the sphere, depending on the materials. Furthermore, we compute the heat emission of a point particle in front of a planar mirror. Finally, we show that a particle placed
inside a spherical mirror cavity does not radiate energy.
\end{abstract}

\pacs{12.20.-m, 
44.40.+a, 
05.70.Ln 
}
\bibliographystyle{plain}

\maketitle

\section{Introduction}
\label{sec:Introduction}
The theory of thermal radiation and radiative heat transfer plays an important role in physics of all length scales: from radiation of the sun to heat transfer between 
nanostructures. Planck's work on blackbody radiation~\cite{Planck1901} together with other revolutionary discoveries led to the birth of quantum theory. Theoretical computations of heat 
radiation (HR) and radiative heat transfer (HT) are based on fluctuational electrodynamics (FE), introduced by Rytov over 60 years
ago~\cite{Rytov1958, Rytov1989}. (It may also be used to study  Casimir interactions~\cite{casimir1948, Dalvit2011}.) 
 The main idea of FE is to relate quantum (thermal) fluctuations of the electromagnetic (EM) field radiated by an electrically neutral object to the fluctuating currents inside 
it. Assuming local equilibrium within the object, the fluctuations can be related to the response function of the object via the fluctuation dissipation 
theorem (FDT)~\cite{Nyquist1928, Callen1951, Weber1955, Kubo1966}.

In the past 50 years, the formalism of FE has been extensively applied to diverse problems of nonequilibrium systems. HR of single macroscopic objects like a semi-infinite plate, a sphere, or an infinitely 
long cylinder was studied in, e.g., Refs.~\cite{Eckhardt1984, Kattawar1970, Bohren2004, Golyk2012}. HT between two plates~\cite{Polder1971, Bimonte2009, Guerout2012}, two spheres~\cite{Narayanaswamy2008}, and 
a sphere and a plate~\cite{Kruger2011, Otey2011, Golyk2013} is particularly interesting due to its dramatic enhancement in the near-field regime. Apart from analytical calculations of nonequilibrium quantities for a particular 
configuration of a couple of simple bodies, there are general theories for arbitrary systems. Numerical scattering techniques were applied to study HT in complex systems with analytically unknown scattering 
properties, e.g., in systems with periodic structures, cones, finite cylinders, or cubes~\cite{Rodriguez2011, McCauley2012, Rodriguez2013, Polimeridis2015}. General formalisms for HR, HT, and nonequilibrium 
Casimir forces in many-body systems have been recently presented and applied~\cite{Kruger2011, Messina2011_1, Messina2011_2, Messina2014, Messina2012, Messina2016, Kruger2012, Muller2017, Bimonte2017}. HT in 
systems, where objects are small compared to all other length scales, can be relatively easily studied numerically, because all the particles can be modeled by dipole 
polarizabilities~\cite{Incardone2014, Ben-Abdallah2011, Nikbakht2014, Dong2017, Wang2016, Choubdar2016, Phan2013}. There are several experimental studies of HT verifying theoretical predictions on both 
qualitative and quantitative levels~\cite{Hargreaves1969, Kittel2005, Shen2009, Rousseau2009, Ottens2011, Kajihara2011, Kim2015}.

Fundamental research in the field of radiative heat transfer is essential for creating various useful technologies. For example, near-field HT is an important mechanism 
for practical applications in energy storage and conversion, thermal management and thermal circuits, near-field imaging and 
nanomanufacturing~\cite{Liu2015, Narayanaswamy2003, Laroche2006, Swanson2009, Ben-Abdallah2014}. Many-body aspects of HT are especially important for implementation 
of devices controlling the heat flux, so-called thermal transistors~\cite{Ben-Abdallah2011, Ben-Abdallah2014}.

In this work, we present a detailed derivation of HR and HT for pointlike nonmagnetic spherical particles (PP)~\footnote{A PP is a nonmagnetic spherical particle which is small compared to any other length scale related to it (e.g., thermal wavelength, skin penetration depth, distances between the particle and other objects).} in the presence of arbitrary objects. Starting from exact expressions 
given in Ref.~\cite{Muller2017} and deriving an explicit form of the scattering operator of a small sphere, we obtain compact, physically insightful formulas for HR and HT. These involve only the Green's function of the surrounding objects and polarizabilities of the particles [see Eqs.~\eqref{eq:HR_pp_final} and \eqref{eq:HT_pp_final} below]. While we take the dipole limit for the  
considered radiating and absorbing particles, the surrounding objects have no restrictions and can be of arbitrary shape, size, and material. As an example, we study the HT between two SiC PPs in the presence of a sphere (SiC, gold, or mirror), where we observe a significant enhancement of the transfer compared to the case where the sphere is absent. This is especially interesting if the sphere is a perfect mirror, as then, the heat transfer from geometric optics vanishes exactly. Furthermore, we analyze the heat emission of a PP in front of a semi-infinite mirror plate. We also demonstrate that a PP placed inside a spherical cavity with perfectly reflecting walls does not radiate 
energy. 

The paper is organized as follows: In Sec.~\ref{sec:FE_general_formulas}, we repeat the exact  expressions for heat radiation and transfer in many-body systems from Ref.~\cite{Muller2017}. These are then used in 
Sec.~\ref{sec:HR_HT_PP_limit} to derive the main formulas of the paper: heat radiation and transfer for point particles. Section~\ref{sec:HR_HT_vac} provides a consistence check of the new formulas, by rederiving known results for isolated 
point particles. In Sec.~\ref{sec:HT_sphere}, we study the HT between point particles in the presence of a macroscopic sphere. Results for the heat emission of a point particle in front of a mirror plate are 
presented in Sec.~\ref{sec:HR_plate}. Section~\ref{sec:HR_cavity} provides our findings for the radiation of a point particle inside a spherical cavity with mirror walls. We close the paper with a summary and discussion in Sec.~\ref{sec:Conclusion}.

\section{Heat radiation and heat transfer in a many-body system}
\label{sec:FE_general_formulas}
In this section, we briefly discuss the setup under study and give general expressions for HR and HT in a many-body system composed of arbitrary objects \cite{Muller2017}. For detailed theories of HR and HT in arbitrary systems, 
we refer the reader to Refs.~\cite{Kruger2011, Messina2011_1, Messina2011_2, Messina2014, Messina2012, Messina2016, Kruger2012, Muller2017}. A description of the relevant EM operators can be found in 
Appendix~\ref{app:EM_operators}.

We consider a system of $ N $ objects labeled by $ \alpha = 1 \dots N $ in vacuum and embedded in an environment. The objects have time-independent homogeneous temperatures 
$ \{T_{\alpha}\} $ and the temperature of the environment is $ T_{\rm env} $. The environment may  be treated as an enclosing blackbody placed sufficiently far away from the
objects. The objects are characterized by their electric and magnetic response, 
$ \bbeps(\omega; \vct{r}, \vct{r}') \equiv \varepsilon_{ij}(\omega; \vct{r}, \vct{r}') $ and 
$ \bbmu(\omega; \vct{r}, \vct{r}') \equiv \mu_{ij}(\omega;\vct{r},\vct{r}') $. 
In the given nonequilibrium (stationary) situation, each object is assumed to be at local equilibrium, such that the current fluctuations within each object independently 
satisfy the FDT~\cite{Nyquist1928, Callen1951, Weber1955, Kubo1966}.

\begin{figure}[!b]
\includegraphics[width=0.8\linewidth]{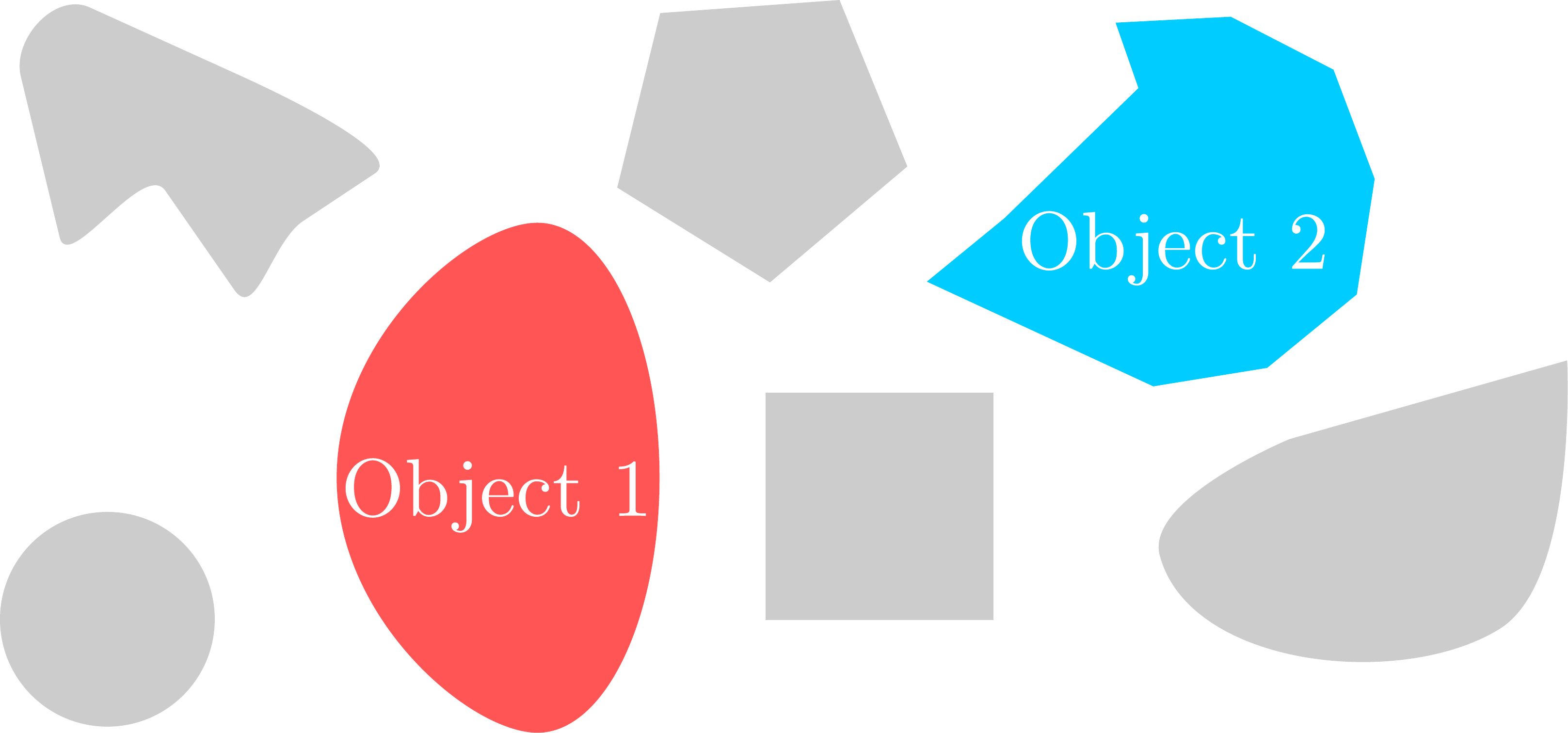}
\caption{\label{fig:system_general}
Configuration of a many-body system in vacuum. Section~\ref{sec:FE_general_formulas} reviews the most general case, where all objects are completely arbitrary, and 
expressions~\eqref{eq:HR_general} and~\eqref{eq:HT_general} can be used to compute HR and HT. In this paper, we focus on the situation, where objects 1 and 2 reduce to spherical point particles, as 
depicted in Figs.~\ref{fig:system_PP_HR} and~\ref{fig:system_PPs_HT}.}
\end{figure}

Our goal is to compute the heat emission $ H_1^{(1)} $ of object 1, i.e., the rate of heat emitted by object 1 and absorbed by it, and the heat transfer rate $ H_1^{(2)} $ from object 1 to object 2, i.e., the 
rate of heat emitted by object 1 and absorbed by object 2 (see Fig.~\ref{fig:system_general}).

$ H_1^{(1)} $ and $ H_1^{(2)} $ read as~\cite{Muller2017}
\begin{widetext}
\begin{align}
H_1^{(1)} =& \frac{2\hbar}{\pi} \int_0^\infty d\omega \frac{\omega}{e^{\frac{\hbar\omega}{k_BT_1}}-1}\Im\Tr\left\{(\mathbb{I}+\mathbb{G}_0\mathbb{T}_{\overline{1}})\frac{1}{\mathbb{I}-\mathbb{G}_0\mathbb{T}_1\mathbb{G}_0\mathbb{T}_{\overline{1}}}\mathbb{G}_0[\Im[\mathbb{T}_1]-\mathbb{T}_1\Im[\mathbb{G}_0]\mathbb{T}_1^*]\frac{1}{\mathbb{I}-\mathbb{G}_0^{ *}\mathbb{T}_{\overline{1}}^*\mathbb{G}_0^*\mathbb{T}_1^*}\right\},
\label{eq:HR_general}\\
\notag H_1^{(2)} =&   \frac{2\hbar}{\pi} \int_0^\infty d\omega \frac{\omega}{e^{\frac{\hbar\omega}{k_BT_1}}-1}\Tr\Bigg\{[\Im[\mathbb{T}_2]-\mathbb{T}_2^*\Im[\mathbb{G}_0]\mathbb{T}_2]\frac{1}{\mathbb{I}-\mathbb{G}_0\mathbb{T}_{\overline{2}}\mathbb{G}_0\mathbb{T}_2}(\mathbb{I}+\mathbb{G}_0\mathbb{T})\frac{1}{\mathbb{I}-\mathbb{G}_0\mathbb{T}_1\mathbb{G}_0\mathbb{T}}\\
&\times\mathbb{G}_0[\Im[\mathbb{T}_1]-\mathbb{T}_1\Im[\mathbb{G}_0]\mathbb{T}_1^*]\frac{1}{\mathbb{I}-\mathbb{G}_0^*\mathbb{T}^*\mathbb{G}_0^*\mathbb{T}_1^*}(\mathbb{I}+\mathbb{G}_0^*\mathbb{T}^*)\mathbb{G}_0^*\frac{1}{\mathbb{I}-\mathbb{T}_2^*\mathbb{G}_0^*\mathbb{T}_{\overline{2}}^*\mathbb{G}_0^*}\Bigg\}.
\label{eq:HT_general}
\end{align}
\end{widetext}

Here, $ c $ is the speed of light in vacuum, $ \hbar $ and $ k_B $ are Planck's and Boltzmann's constants, respectively; $ \mathbb{T}_1 $ and $ \mathbb{T}_2 $ are the scattering operators of object 1 and 2, respectively; 
$ \mathbb{T}_{\overline{i}} $ is the operator of the system without object $i$ present, and $ \mathbb{T} $ is the composite operator of the gray objects in Fig.~\ref{fig:system_general} (i.e., of  the system with objects 1 and 2 absent);  $ \mathbb{G}_0 $ is the free Green's function (GF)~\footnote{In this paper, we use words \enquote{Green's function} and symbol $ \mathbb{G} $ for \textit{electric} Green's dyad.}; $ \mathbb{I} $ is the identity 
operator. We emphasize that Eqs.~\eqref{eq:HR_general} and~\eqref{eq:HT_general} contain no approximations regarding the properties of the objects.

Using the expressions for HR and HT, one can compute other important nonequilibrium quantities. For example, the \textit{total} heat absorbed by object 1, i.e., the heat radiated by all objects 
and the environment and absorbed by object 1, is given by the sum over the heat transfer contributions from all objects~\cite{Kruger2012, Muller2017}:
\begin{equation}
H^{(1)}(\{T_\alpha\},T_{\rm env})=\sum_{\alpha=1}^{N} \left[H_\alpha^{(1)}(T_\alpha)- H_\alpha^{(1)}(T_{\rm env})\right].
\label{eq:total_absorption}
\end{equation}
Another important quantity is the \textit{net} HT from object 1 to object 2, i.e., the heat emitted by object 1 and absorbed by object 2 minus the heat emitted by object 2 and absorbed by object 1:
\begin{equation}
H^{1\rightarrow2}=H_1^{(2)}(T_1)-H_2^{(1)}(T_2)=H_1^{(2)}(T_1)-H_1^{(2)}(T_2),
\label{eq:total_HT}
\end{equation}
where the last step follows from the symmetry of HT~\footnote{The symmetry of HT means that $ H_{1}^{(2)}(T)=H_{2}^{(1)}(T) $, which implies no \textit{net} HT  $ H^{1\rightarrow2}=H_1^{(2)}(T_1)-H_2^{(1)}(T_2) $ between the objects at equal temperatures. Using the positivity and the symmetry of HT, one can show that the net energy flow is always from the warmer object to the colder one~\cite{Kruger2012, Muller2017}.}. Expressions~\eqref{eq:total_absorption} and~\eqref{eq:total_HT} reflect the principle of detailed balance: at global 
thermal equilibrium, where all temperatures are equal, all radiative fluxes cancel each other, such that $ H^{(1)} = H^{1\rightarrow2} = 0 $.

\section{Heat radiation and transfer for pointlike spherical particles in arbitrary geometries}
\label{sec:HR_HT_PP_limit}
In this section, we first deduce the scattering operator of a small sphere. By evaluating Eqs.~\eqref{eq:HR_general} and~\eqref{eq:HT_general} in the PP limit for objects 1 and 2, we derive formulas 
for HR and HT for PPs in the presence of an arbitrary collection of objects.

\subsection{Scattering operator of a small sphere}
Consider a homogeneous isotropic nonmagnetic ($ \mu $ = 1) sphere of radius $R$ with local potential (see Appendix~\ref{app:EM_operators} for its relation to the Helmholtz equation)
\begin{equation}
\mathbb{V}_s(\vct{r}, \vct{r}') = 
\begin{cases}
k^2(\varepsilon-1)\mathbb{I}(\vct{r}, \vct{r}'), & \vct{r}, \vct{r}' \in V_s\\
0, & {\rm else}.
\end{cases}
\label{eq:potential_s}
\end{equation}
where $ k = \frac{\omega}{c} $, $ \varepsilon $ is the frequency-dependent scalar dielectric function, $ \mathbb{I}(\vct{r}, \vct{r}') $ is the identity operator, and $ V_s $ is the 
volume of the sphere. 
The scattering operator of the sphere [the quantity appearing in Eqs.~\eqref{eq:HR_general} and~\eqref{eq:HT_general}] is defined as (see Appendix~\ref{app:EM_operators})
\begin{equation}
\mathbb{T}_s = \mathbb{V}_s\frac{1}{\mathbb{I}-\mathbb{G}_0\mathbb{V}_s}.
\label{eq:T_s_def}
\end{equation}
Substituting Eq.~\eqref{eq:potential_s} into Eq.~\eqref{eq:T_s_def}, we have more explicitly 
\begin{equation}
\mathbb{T}_s(\vct{r}, \vct{r}') = 
\begin{cases}
k^2(\varepsilon-1)[\mathbb{I}-\mathbb{G}_0\mathbb{V}_s]^{-1}(\vct{r}, \vct{r}'), & \vct{r}, \vct{r}' \in V_s\\
0, & {\rm else}.
\end{cases}
\label{eq:T_s}
\end{equation}
Evaluating the inverse operator $ [\mathbb{I}-\mathbb{G}_0\mathbb{V}_s]^{-1} $ is in general challenging, and Eqs.~\eqref{eq:HR_general} and~\eqref{eq:HT_general} are typically evaluated using partial waves \cite{Kruger2012,Muller2017}.

Great simplification, however, can be made if the sphere is sufficiently small, so that the EM wave does not feel the internal structure of the sphere, i.e., the electric field within the sphere is uniform in space and time. This is the case if, for a frequency  $ \omega $ and the corresponding wavelength $ \lambda $, one has~\cite{Bohren2004}
\begin{equation}
R \ll \lambda, \hspace{1cm} R \ll \frac{\lambda}{2\pi|\sqrt{\varepsilon(\omega)}|}.
\label{eq:small_sphere_conditions}
\end{equation}
This limit is the electrostatic dipole limit for the scattering of the field by a sphere~\cite{Bohren2004}. When using this limit in the  integration over frequency [e.g., in Eqs.~\eqref{eq:HR_general} and~\eqref{eq:HT_general}], the condition in Eq.~\eqref{eq:small_sphere_conditions} should be fulfilled by frequencies which contribute dominantly, which are usually around $ \omega_T = \frac{2\pi k_BT}{\hbar} $ (with the corresponding thermal wavelength $ \lambda_T = \frac{\hbar c}{k_BT} $). 

In the limit of Eq.~\eqref{eq:small_sphere_conditions}, the field inside the sphere takes a simple form \cite{Bohren2004, Tsang2000}. Using the Lippmann-Schwinger equation~\cite{Kruger2012, Lippmann1950} as well as the relation between incident field and the field inside the sphere, the operator in Eq.~\eqref{eq:T_s} can be found (here, $ \vct{r}, \vct{r}'$ are inside the sphere):
\begin{align}
\notag [\mathbb{I}-\mathbb{G}_0\mathbb{V}_s]_{ss}(\vct{r}, \vct{r}') &= \mathbb{I}(\vct{r}, \vct{r}') + \frac{1}{3k^2}\mathbb{V}_s(\vct{r}, \vct{r}')\\
& = \frac{\varepsilon+2}{3}\mathbb{I}(\vct{r}, \vct{r}').
\label{eq:inverse_operator_in_T}
\end{align}
Because the identity is its own inverse, we have from Eq.~\eqref{eq:T_s} a closed form for the scattering operator of a small sphere,
\begin{equation}
\mathbb{T}_{ss}(\vct{r}, \vct{r}') = 
\begin{cases}
3k^2\frac{\varepsilon-1}{\varepsilon+2}\mathbb{I}(\vct{r}, \vct{r}'), & \vct{r}, \vct{r}' \in V_{ss}\\
0, & {\rm else}.
\end{cases}
\label{eq:T_ss}
\end{equation} 
We have used indices $ss$ to label the small sphere.

The scattering operator of a small sphere is proportional to the identity operator, i.e., it is diagonal and local, which will simplify the expressions for HR and HT. The prefactor is identified with the electrical dipole polarizability 
\begin{equation}
\alpha = \frac{\varepsilon-1}{\varepsilon+2}R^3
\label{eq:polarazability}
\end{equation}
of a homogeneous isotropic sphere with radius $ R $. To check formula~\eqref{eq:T_ss}, we show in 
Appendix~\ref{app:ss_operator_matrix_correspondence} that the operator $ \mathbb{T}_{ss}(\vct{r}, \vct{r}') $ yields the correct scattering matrix elements in the small sphere limit.

\subsection{Simplifying Eqs.~\eqref{eq:HR_general} and~\eqref{eq:HT_general} for point particles}
\begin{figure}
\includegraphics[width=0.8\linewidth]{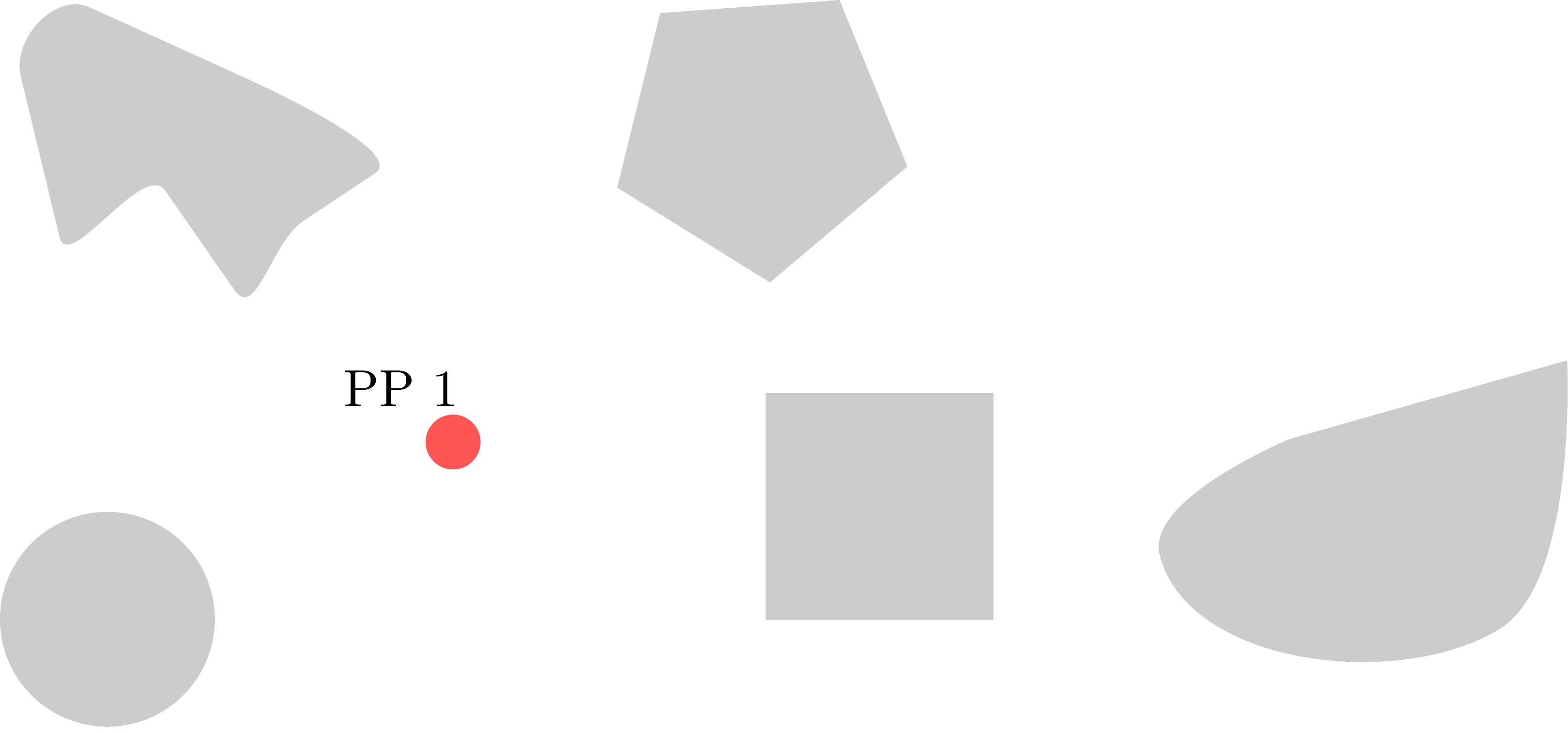}
\caption{\label{fig:system_PP_HR}
System where object 1 is a PP. One can use formula~\eqref{eq:HR_pp_final} to compute the HR of this PP in this system.}
\end{figure}

\begin{figure}
\includegraphics[width=0.8\linewidth]{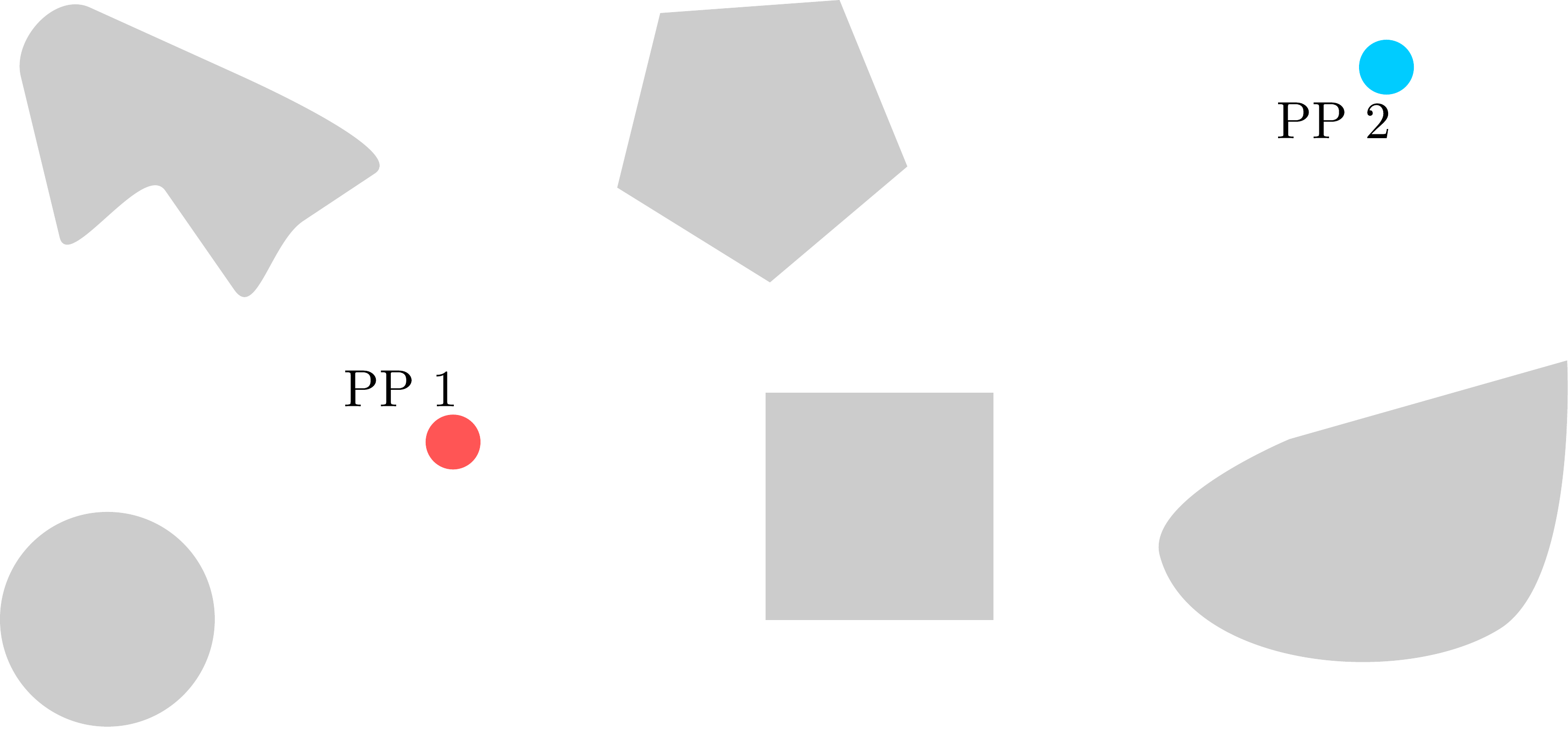}
\caption{\label{fig:system_PPs_HT}
System where objects 1  and 2 are PPs. One can use formula~\eqref{eq:HT_pp_final} to compute HT from PP 1 to PP 2 in this system.}
\end{figure}

Additionally to the condition \eqref{eq:small_sphere_conditions}, let the spheres be small compared to their distance to any other object. In other words, starting from  the setup of Fig.~\ref{fig:system_general}, we let objects 1 and 2 become small (pointlike) particles as in Figs.~\ref{fig:system_PP_HR} and~\ref{fig:system_PPs_HT}.
Note that object~2 in Fig.~\ref{fig:system_general} is not present in Fig.~\ref{fig:system_PP_HR}, because it has no special role for the heat radiation of object 1 in Eq.~\eqref{eq:HR_pp_initial} below. One may imagine that it was removed, or that it became part of the gray objects.

In this point-particle limit we can neglect multiple scatterings and use a one-reflection approximation between the PPs and the residual objects~\cite{Kruger2012}, i.e., the inverse operators in Eqs.~\eqref{eq:HR_general} and \eqref{eq:HT_general} are set to unity. 
Furthermore,  we only keep the terms linear in the scattering operator of the PPs (because the quadratic terms are negligible in this limit).


\subsection{Heat radiation of a point particle}
Evaluation of Eq.~\eqref{eq:HR_general} in the point-particle limit reads as (we set $\mathbb{T}_2 = 0 $ in Eq.~\eqref{eq:HR_general}, see Fig.~\ref{fig:system_PP_HR})
\begin{equation}
H_{1pp}^{(1pp)} = \frac{2\hbar}{\pi} \int_0^\infty d\omega \frac{\omega}{e^{\frac{\hbar\omega}{k_BT_1}}-1}\Tr\{\Im[\mathbb{G}]\Im[\mathbb{T}_{1ss}]\},
\label{eq:HR_pp_initial}
\end{equation}
where $ \mathbb{G} = \mathbb{G}_0 + \mathbb{G}_0\mathbb{T}\mathbb{G}_0 $ (see Appendix~\ref{app:EM_operators}) is, as in Eq.~\eqref{eq:HR_general}, the GF of the system in the absence of PP 1 (the gray objects 
in Fig.~\ref{fig:system_PP_HR}), and $ \mathbb{T}_{1ss} $ is the scattering operator of the particle given in Eq.~\eqref{eq:T_ss}. Using Eq.~\eqref{eq:T_ss}, we have
\begin{align}
\notag H_{1pp}^{(1pp)} =& \ \frac{6\hbar}{\pi c^2} \int_0^\infty d\omega \frac{\omega^3}{e^{\frac{\hbar\omega}{k_BT_1}}-1} \Im \left[ \frac{\varepsilon_1-1}{\varepsilon_1+2}\right]\\ 
& \times \int_{V_1} d^3r \sum_i \Im G_{ii}(\vct{r}, \vct{r}),
\label{eq:HR_pp_f1}
\end{align}
where $ V_1 $ is the volume of the particle. Note that both $ \varepsilon_1 $ and $ G_{ii}(\vct{r}, \vct{r}) $ are frequency-dependent quantities. Because the particle, PP 1, is small compared to its distance to other objects, the value of $G_{ii}(\vct{r}, \vct{r})$ hardly varies between different points inside PP 1. In this limit, the integral in Eq.~\eqref{eq:HR_pp_f1} thus yields a factor of the particle's volume, and the argument $ \vct{r} $  is replaced by $ \vct{r}_1 $, the position of the PP. We finally obtain,
\begin{equation}
H_{1pp}^{(1pp)} = \frac{8\hbar}{c^2} \int_0^\infty d\omega \frac{\omega^3}{e^{\frac{\hbar\omega}{k_BT_1}}-1} \Im (\alpha_1) \sum_i \Im G_{ii}(\vct{r}_1, \vct{r}_1),
\label{eq:HR_pp_final}
\end{equation}
where we introduced the electrical dipole polarizability $ \alpha_1 $ defined in Eq.~\eqref{eq:polarazability}. Equation~\eqref{eq:HR_pp_final} is our first main result.

The term $ \sum_i \Im G_{ii}(\vct{r}_1, \vct{r}_1) $ in Eq.~\eqref{eq:HR_pp_final} is identified with the electric part of the local EM density of states~\cite{Joulain2003, Joulain2005}. Note that the HR is proportional to the volume of the particle, a feature that is inherent to small objects~\cite{Kruger2012, Golyk2012}. We also note that the quantity $ H_{1pp}^{(1pp)} $ is non-negative as is the more general expression $H_1^{(1)}$ in Eq.~\eqref{eq:HR_general} \cite{Muller2017}.

\subsection{Heat transfer between two point particles}
Let us turn to the heat transfer between PP~1 and PP~2 in Fig.~\ref{fig:system_PPs_HT}. Equation~\eqref{eq:HT_general} evaluated in the point-particle limit reads
\begin{align}
\notag H_{1pp}^{(2pp)} =&  \frac{2\hbar}{\pi} \int_0^\infty d\omega \frac{\omega}{e^{\frac{\hbar\omega}{k_BT_1}}-1}\\
& \times \Tr\{\Im[\mathbb{T}_{2ss}]\mathbb{G}\Im[\mathbb{T}_{1ss}]\mathbb{G}^*\},
\label{eq:HT_pp_initial}
\end{align}
where, again, $ \mathbb{G} = \mathbb{G}_0 + \mathbb{G}_0\mathbb{T}\mathbb{G}_0 $ (see Appendix~\ref{app:EM_operators}) is the Green's function of the system in the absence of the two point particles (the system composed of gray objects in Fig.~\ref{fig:system_PPs_HT}). $ \mathbb{T}_{1ss} $ ($ \mathbb{T}_{2ss} $) is the scattering operator of PP 1  (PP 2). Using expression~\eqref{eq:T_ss} 
for the scattering operator of a small sphere, the trace can be readily performed,
\begin{align}
\notag \Tr&\{\Im[\mathbb{T}_{2ss}]\mathbb{G}\Im[\mathbb{T}_{1ss}]\mathbb{G}^*\}\\
\notag & =  9k^4 \Im \left[ \frac{\varepsilon_1-1}{\varepsilon_1+2}\right] \Im \left[ \frac{\varepsilon_2-1}{\varepsilon_2+2}\right]\\
& \ \ \ \times\int_{V_2} d^3r \int_{V_1} d^3r' \sum_{ij}|G_{ij}(\vct{r}, \vct{r}')|^2,
\label{eq:HT_pp_tr}
\end{align}
where $ V_1 $ and $ V_2 $ are volumes of the particles. In the above equation we used reciprocity, $ G_{ij}(\vct{r}, \vct{r}')=G_{ji}(\vct{r}', \vct{r}) $~\cite{Eckhardt1984}.  
Also here, we can assume that the value of $G_{ij}(\vct{r}, \vct{r}')$ hardly varies for different points inside the two PPs, and we finally obtain for the HT
\begin{widetext}
\begin{equation}
H_{1pp}^{(2pp)} = \frac{32\pi\hbar}{c^4} \int_0^\infty d\omega \frac{\omega^5}{e^{\frac{\hbar\omega}{k_BT_1}}-1}\Im(\alpha_1)\Im(\alpha_2)\sum_{ij}|G_{ij}(\vct{r}_2, \vct{r}_1)|^2,
\label{eq:HT_pp_final}
\end{equation}
\end{widetext}
where $ \vct{r}_1 $ and $ \vct{r}_2 $ are the coordinates of the particles. 
Note that the HT given by Eq.~\eqref{eq:HT_pp_final} is positive and symmetric, and it is proportional to the particles' volumes.

Equations~\eqref{eq:HR_pp_final} and~\eqref{eq:HT_pp_final} are the main results of this paper. They give HR of a PP and HT between two PPs in the presence of an arbitrary system of objects 
(see Figs.~\ref{fig:system_PP_HR} and~\ref{fig:system_PPs_HT}). We emphasize that the formulas imply no simplifications for the objects surrounding the two particles. Note that the GF in 
formulas~\eqref{eq:HR_pp_final} and~\eqref{eq:HT_pp_final} does not include the two PPs. Thus, in order to study HR of a PP and HT between two PPs, one has to know the GF of 
the surrounding objects. The evaluation of the derived formulas for different example geometries is given in the subsequent sections. 

Studying the case of a collection of small spherical particles, a formula similar to Eq.~\eqref{eq:HT_pp_final} was found in Ref.~\cite{Ben-Abdallah2011} (however, in terms of the GF
including the radiating and absorbing particles).

\subsection{Radiation and transfer for small particles of arbitrary shape}
While Eqs.~\eqref{eq:HR_pp_final} and~\eqref{eq:HT_pp_final} are valid for spherical point particles, the more general formulas~\eqref{eq:HR_pp_initial} and~\eqref{eq:HT_pp_initial} are valid for particles of arbitrary shape. The heat emission of a particle of arbitrary shape with scattering operator $\mathbb{T}_1$ is thus
\begin{equation}
H_{1pp}^{(1pp)} = \frac{2\hbar}{\pi} \int_0^\infty d\omega \frac{\omega}{e^{\frac{\hbar\omega}{k_BT_1}}-1}\Tr\{\Im[\mathbb{G}]\Im[\mathbb{T}_{1}]\}.
\label{eq:HR_a}
\end{equation}
The heat transfer between particles 1 and 2, both of arbitrary shapes,  with scattering operators $\mathbb{T}_1$ and $\mathbb{T}_2$ is 
\begin{align}
 H_{1pp}^{(2pp)} =&  \frac{2\hbar}{\pi} \int_0^\infty d\omega \frac{\omega}{e^{\frac{\hbar\omega}{k_BT_1}}-1}
  \Tr\{\Im[\mathbb{T}_{2}]\mathbb{G}\Im[\mathbb{T}_{1}]\mathbb{G}^*\}.
\label{eq:HT_a}
\end{align}
Eqs.~\eqref{eq:HR_a} and \eqref{eq:HT_a} have the same regime of validity as Eqs.~\eqref{eq:HR_pp_initial} and~\eqref{eq:HT_pp_initial} regarding the size of the particles, their material properties, and distances to other objects.

\section{Rederivation of heat radiation and heat transfer in vacuum}
\label{sec:HR_HT_vac}
The results for HR of a small sphere and HT between PPs without additional objects present are well known in literature~\cite{Kruger2012, Volokitin2001, Chapuis2008}, and serve as a first test of Eqs.~\eqref{eq:HR_pp_final} and \eqref{eq:HT_pp_final}  obtained in the previous section. In this case, the GF appearing  in Eqs.~\eqref{eq:HR_pp_final} and~\eqref{eq:HT_pp_final} is the free GF given in Eq.~\eqref{eq:GF_free}. It is straightforward to show that
\begin{align}
& \sum_i \Im G_{0ii}(\vct{r}_1, \vct{r}_1) = \frac{k}{2\pi},\label{eq:trace_ImG0_vac}\\
& \sum_{ij}|G_{0ij}(\vct{r}_2, \vct{r}_1)|^2 = \frac{1}{8\pi^2 d^2}\left[1+\frac{1}{k^2d^2}+\frac{3}{k^4d^4}\right],\label{eq:trace_AbsSqG0_vac}
\end{align}
where $ d = |\vct{r}_2 - \vct{r}_1| $ is the distance between the particles [kept finite in Eq.~\eqref{eq:trace_AbsSqG0_vac}]. Substituting expression~\eqref{eq:trace_ImG0_vac} into formula~\eqref{eq:HR_pp_final} we find for the HR of a PP in vacuum
\begin{equation}
H_{1pp, vac}^{(1pp)} = \frac{4\hbar}{\pi c^3} \int_0^\infty d\omega \frac{\omega^4}{e^{\frac{\hbar\omega}{k_BT_1}}-1} \Im (\alpha_1),
\label{eq:HR_pp_vac}
\end{equation}
which reproduces Eq.~(129) in Ref.~\cite{Kruger2012}. Note that in vacuum a PP and a small sphere are conceptually identical, and the above expression is thus valid if the conditions~\eqref{eq:small_sphere_conditions} 
are fulfilled. Substitution of expression ~\eqref{eq:trace_AbsSqG0_vac} into formula~\eqref{eq:HT_pp_final} gives the HT between PPs in vacuum
\begin{align}
\notag H_{1pp, vac}^{(2pp)} =& \ \frac{4\hbar}{\pi c^4} \int_0^\infty d\omega \frac{\omega^5}{e^{\frac{\hbar\omega}{k_BT_1}}-1}\Im(\alpha_1)\Im(\alpha_2)\\ 
& \times \left[\frac{1}{d^2}+\frac{c^2}{\omega^2d^4}+\frac{3c^4}{\omega^4d^6}\right],
\label{eq:HT_pp_vac}
\end{align}
which is in agreement with Eq.~(137) in Ref.~\cite{Kruger2012}, Eq.~(62) in Ref.~\cite{Volokitin2001} and Eq.~(9) in Ref.~\cite{Chapuis2008}.

\section{Heat transfer in the presence of a sphere of arbitrary size}
\label{sec:HT_sphere}
\begin{figure}[b]
\includegraphics[width=0.8\linewidth]{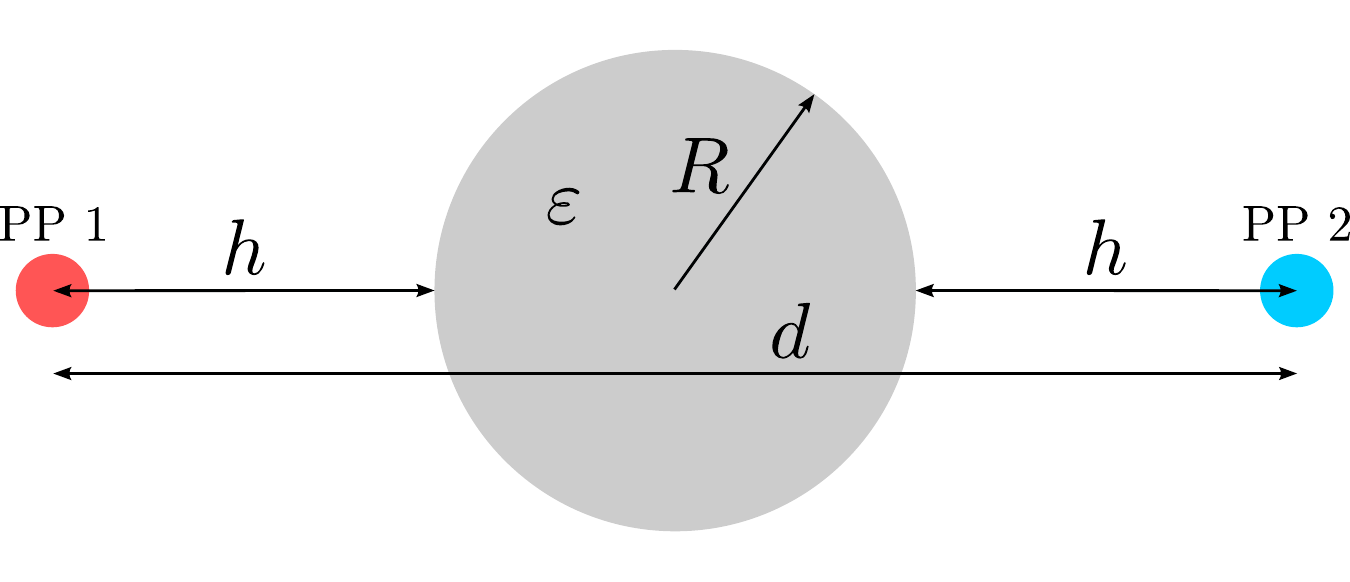}
\caption{\label{fig:config_PPs_sphere}Configuration studied in Fig.~\ref{fig:HT_PPs_sphere}.  HT between PPs in the presence of a sphere.}
\end{figure}
\begin{figure*}
\includegraphics[width=1.0\linewidth]{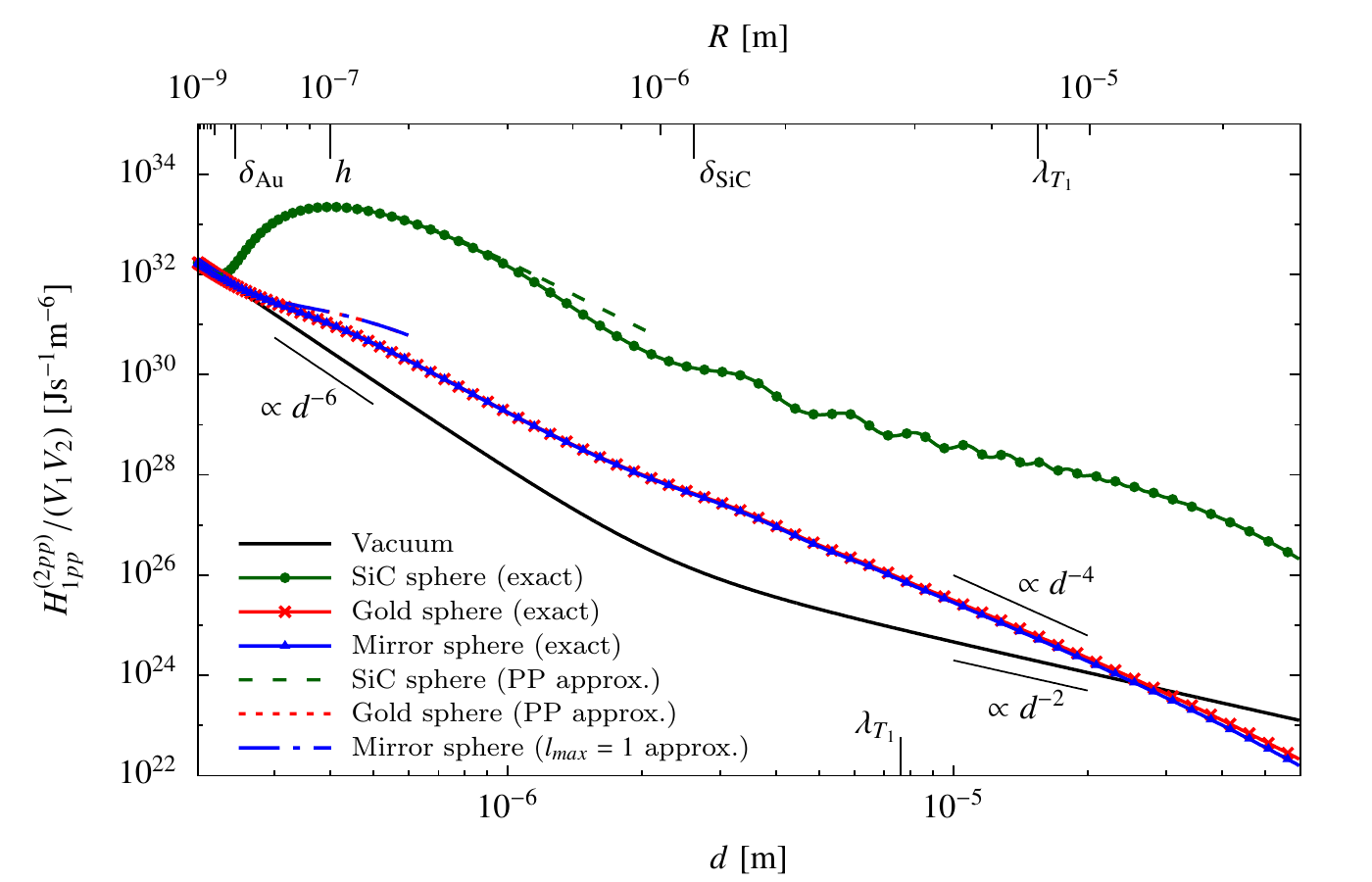}
\caption{\label{fig:HT_PPs_sphere}Normalized (by the PPs volumes) heat transfer from SiC PP 1 at $ T_1 = 300 \ {\rm K } $ to SiC PP 2 in the presence of a sphere, as a function of distance (lower axis) between the particles (see Fig.~\ref{fig:config_PPs_sphere}). Upper $x$ axis shows the corresponding sphere's radius. The solid black curve is the result without the sphere present [Eq.~\eqref{eq:HT_pp_vac}]. The solid curves with points correspond to the HT in the presence of the sphere, with material as labeled. Labels \enquote{PP approximation} give the HT where the  sphere is approximated by the GF of a PP given in formula~\eqref{eq:GF_pp}. For the mirror sphere, we show the approximation, where the sum in the GF [Eq.~\eqref{eq:GF_sphere}] was restricted 
to $ l = 1 $. On the lower axis, we give the thermal wavelength $\lambda_{T_1}$, while on the upper axis, we also give $h$, and the skin depths $\delta$ of gold and SiC.}  
\end{figure*}

In this section, we study the HT between two PPs in the presence of a homogeneous isotropic nonmagnetic ($ \mu = 1 $) sphere of radius $R$ (see Appendix~\ref{app:GFs} for the GF of a sphere, which is given as a sum over multipoles). 
We do not make any approximations regarding the size of the sphere and use as many multipoles as needed for the convergence of the sum in the GF.  

In particular, we consider the configuration depicted in Fig.~\ref{fig:config_PPs_sphere}: The PPs are placed symmetrically at fixed distance $ h = 10^{-7} \ {\rm m} $ at opposite sides of the sphere's surface, so that their mutual distance is $ d = 2(R+h) $. The radius $ R $ is varied from $ 10^{-9} \ {\rm m} $ up to $ 3\times10^{-5} \ {\rm m} $. We evaluate expression~\eqref{eq:HT_pp_final} with temperature $ T_1 = 300 \ {\rm K} $, and let the PPs be made of SiC, using the following dielectric 
function \cite{Spitzer1959}:
\begin{equation}
\varepsilon_{\rm SiC}(\omega) = \varepsilon_\infty\frac{\omega^2-\omega_{\rm LO}^2+i\omega\gamma}{\omega^2-\omega_{\rm TO}^2+i\omega\gamma},
\label{eq:epsilon_SiC}
\end{equation}
where  $ \varepsilon_\infty=6.7 $, $ \omega_{\rm LO}=1.82\times10^{14} \ {\rm rad} \ {\rm s}^{-1} $, $ \omega_{\rm TO}=1.48\times10^{14} \ {\rm rad} \ {\rm s}^{-1} $, 
$ \gamma=8.93\times10^{11} \ {\rm rad} \ {\rm s}^{-1} $. As regards the sphere, we consider three different materials: SiC, perfect mirror, and gold, for which the Drude model was used,
\begin{equation}
\varepsilon_{\rm Au}(\omega) = 1-\frac{\omega_p^2}{\omega(\omega+i\omega_{\tau})},
\label{eq:epsilon_gold}
\end{equation}
with $ \omega_p = 1.37\times 10^{16} \ {\rm rad} \ {\rm s}^{-1} $ and $ \omega_{\tau} = 4.06\times 10^{13} \ {\rm rad} \ {\rm s}^{-1} $.
Note that, in addition to conditions~\eqref{eq:small_sphere_conditions}, the size of the particles must be much smaller than $ h $ for the PP limit to be valid. Because we normalize the HT by the volumes of the PPs, we do not give their sizes explicitly. 

Figure~\ref{fig:HT_PPs_sphere} shows the resulting HT, where in general a significant enhancement of the HT due to the presence of the sphere for a large range of $R$ is visible. When the radius is small 
compared to $h$, the HT approaches the vacuum result, because the presence of the sphere becomes less and less relevant; similar behavior was observed in Refs.~\cite{Ben-Abdallah2011, Dong2017}. Once the radius becomes comparable to the distance $ h $, the HT starts to deviate from the vacuum result. 

The effect is strongest for SiC, for which the curve has a local maximum  at $ R \approx 10^{-7} \ {\rm m} $, i.e., when $ R \approx h $. For larger $R$ the HT decreases, and shows oscillations in the range  $ R \approx \delta_{\rm SiC} $. Overall, the SiC sphere gives an enhancement of around four orders of magnitude for almost the whole range of radii shown.

A very different physical setup is given by the other two materials. (Because gold and the perfect mirror show almost identical results, we will not distinguish between them in our discussion). Here, the waves cannot penetrate the sphere, and a naive estimate of HT from a view factor~\cite{Modest2013} would yield exactly zero. Nevertheless, due to diffraction effects, the HT is finite, and is even  significantly larger compared to the absence of the sphere for $R \lessapprox \lambda_{T_1} $. The diffraction seems to have a lensing effect. For $R \gg \lambda_{T_1} $, diffraction is absent, and the HT goes to zero faster than the vacuum result.  Between these limits, 
there is an intersection with the vacuum HT, a \enquote{cloaking point}. Here, the big sphere does not affect the HT and is hence \enquote{invisible} from the viewpoint of HT. 

Partly related setups were studied in Refs.~\cite{Ben-Abdallah2011, Dong2017}, where, however, different limits were investigated, namely, the case where  \textit{all} particles are much 
smaller than the thermal wavelength, in contrast to Fig.~\ref{fig:HT_PPs_sphere}. Also, the variation of parameters is different. While the radius of the particles in 
Refs.~\cite{Ben-Abdallah2011, Dong2017} was fixed and the distance between them was varied, we change the radius of the intermediate sphere keeping the distance from the particles to the sphere's surface constant. 
Therefore, our results differ from those presented in Refs.~\cite{Ben-Abdallah2011, Dong2017}. 

\begin{figure}
\includegraphics[width=1.0\linewidth]{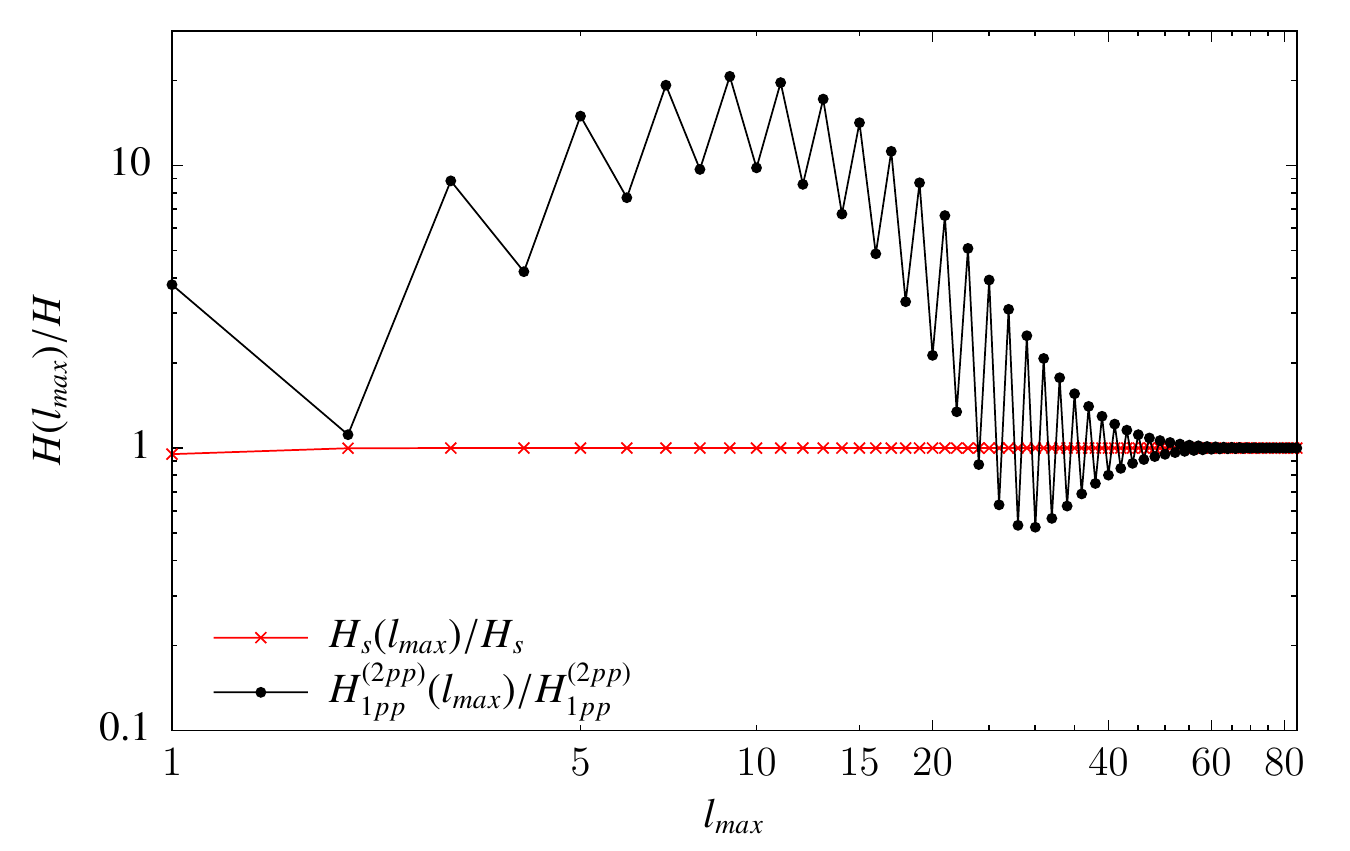}
\caption{\label{fig:GF_sphere_conv}Convergence of the HT between SiC PPs in the presence of a gold sphere of radius $ R = 10^{-6} \ {\rm m} $ as a function of  
 the maximum  multipole order used in the sum, normalized by the exact value (other parameters as in Fig.~\ref{fig:HT_PPs_sphere}). For comparison, also the corresponding curve for the heat radiation of an isolated gold sphere of radius $ R = 10^{-6} \ {\rm m} $  is shown. Connecting lines are included as a guide to the eye.}
\end{figure}

We would like to mention that the numerical convergence of the shown results, as regards summation over multipoles, is nontrivial, as shown in Fig.~\ref{fig:GF_sphere_conv}. 
The figure shows the result for gold and $R= 10^{-6} \ {\rm m} $ of Fig.~\ref{fig:HT_PPs_sphere}, truncated at multipole order $l_{max}$ normalized by the exact value ($l_{max}\to \infty$). It is visible, that the curve oscillates, and approaches the final result for $l_{max}\approx 60$ only. In comparison, the heat radiation of the same sphere in isolation converges much faster (also shown, using Eq.~(124) in Ref.~\cite{Kruger2012}). The difference is due to the small length $h$ present in the three-body configuration. For the points with the largest $ R $ shown in Fig.~\ref{fig:HT_PPs_sphere}, $l_{max}$ was of the order of several thousands.

Lastly, extracting from Fig.~\ref{fig:HT_PPs_sphere}, the sphere can be modeled by a PP if $R$ is small compared to $h$ and the skin depth $\delta$. For SiC, this approximation appears valid up to  even slightly larger values of $R$.

\section{Heat radiation in the presence of a mirror plate}
\label{sec:HR_plate}
\begin{figure}[t]
\includegraphics[width=0.8\linewidth]{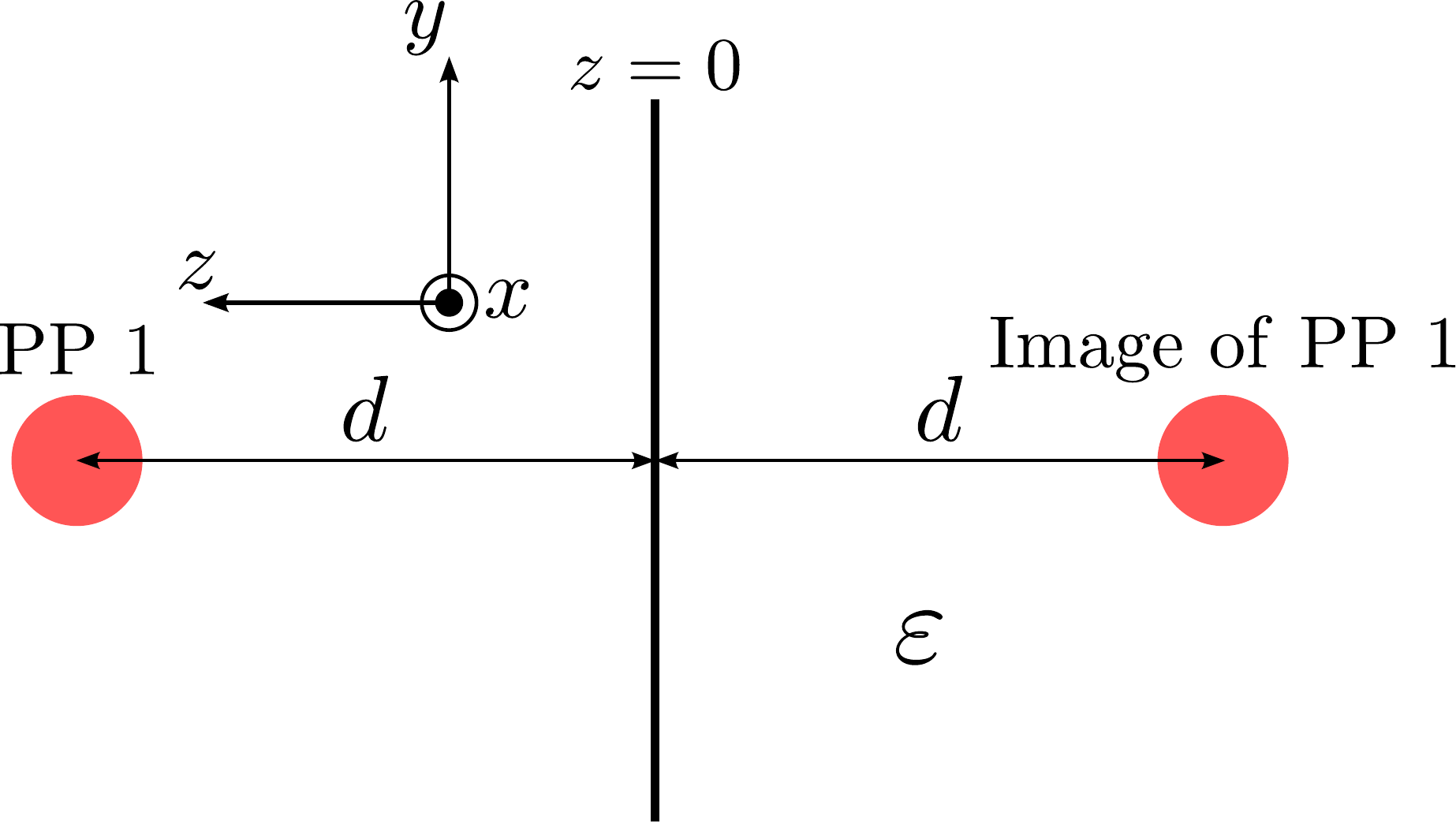}
\caption{\label{fig:config_PP_plate}PP 1 in front of a plate with the dielectric function $ \varepsilon $ occupying the space $ z \leqslant 0 $. While Appendix~\ref{app:GFs} gives the case of a general $\varepsilon$, in the main text, we focus  on $ |\varepsilon| \to \infty $ (perfect reflector). It is thus the HR of the particle that sees itself in the mirror.}
\end{figure}

In this section, we study the HR of a PP in the presence of a mirror plate at distance $d$ (see Fig.~\ref{fig:config_PP_plate}). 
In electrostatics, this problem is addressed with the method of images~\cite{Jackson1999, Barnes1998}. 
The GF of a plate is well known and given in Appendix~\ref{app:GFs}. 
The following derivation (arriving at Eq.~\eqref{eq:HR_mirror_plate} below) is valid if conditions \eqref{eq:small_sphere_conditions} are fulfilled and the distance $ d $ is large compared to the particle's radius.  We thus  use formula~\eqref{eq:HR_pp_final} to evaluate the HR.
We need only the trace of the imaginary part of the GF, where both arguments are equal to the position $ \vct{r}_1 $ of the PP. Substituting plane waves~\eqref{eq:plane_waves_M},~\eqref{eq:plane_waves_N} and Fresnel coefficients~\eqref{eq:Fresnel_coeff_M_mirror},~\eqref{eq:Fresnel_coeff_N_mirror} into GF~\eqref{eq:GF_plate}, one finds 
\begin{align}
\notag \sum_i & \Im G_{ii}(\vct{r}_1, \vct{r}_1) = \sum_i \Im G_{0ii}(\vct{r}_1, \vct{r}_1)\\
& - \Im \left\{\frac{i}{4\pi^2}\int d^2k_\perp\frac{1}{k^2}\sqrt{k^2-k_\perp^2}e^{2id\sqrt{k^2-k_\perp^2}}\right\}.
\label{eq:G_mirror_plate_trace}
\end{align}
The first term in the above expression, the vacuum term, is given in Eq.~\eqref{eq:trace_ImG0_vac}. The integral in the second term can be performed, and we obtain
\begin{align}
\notag \sum_i & \Im G_{ii}(\vct{r}_1, \vct{r}_1) = \frac{k}{2\pi} - \frac{1}{8\pi k^2 d^3}[-\sin{(2kd)}\\
& +2kd\cos{(2kd)}+2k^2d^2\sin{(2kd)}].
\label{eq:ImG_mirror_plate_trace_final}
\end{align}
The first term gives the HR of the particle in isolation. Substituting the above expression into formula~\eqref{eq:HR_pp_final}, we finally have for the radiation of a PP in front of a mirror plate
\begin{widetext}
\begin{equation}
H_{1pp}^{(1pp)} = H_{1pp, vac}^{(1pp)} - \frac{\hbar}{\pi d^3} \int_0^\infty d\omega \frac{\omega}{e^{\frac{\hbar\omega}{k_BT_1}}-1} \Im (\alpha_1)\left[-\sin{\left(2\frac{\omega}{c}d\right)}+2\frac{\omega}{c}d\cos{\left(2\frac{\omega}{c}d\right)}+2\frac{\omega^2}{c^2}d^2\sin{\left(2\frac{\omega}{c}d\right)}\right],
\label{eq:HR_mirror_plate}
\end{equation}
\end{widetext}
where $ H_{1pp, vac}^{(1pp)} $ is the HR of the PP in vacuum given by formula~\eqref{eq:HR_pp_vac}.

\begin{figure}[b]
\includegraphics[width=1.0\linewidth]{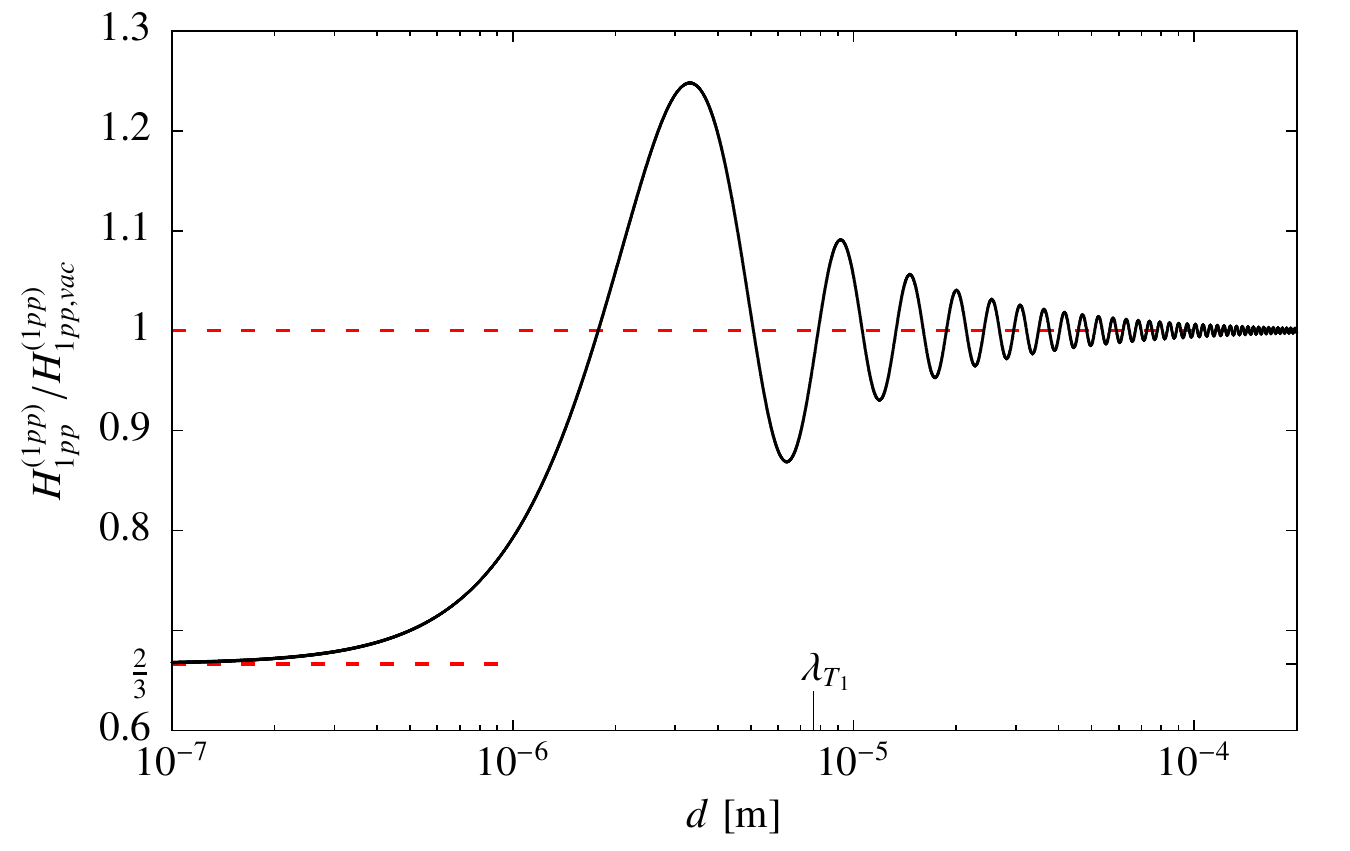}
\caption{\label{fig:HR_PP_plate} HR of a SiC PP 1 at temperature $ T_1 = 300 \ {\rm K} $ in front of a mirror plate as a function of distance from the plate, normalized by the result of an isolated particle, from Eq.~\eqref{eq:HR_mirror_plate}. Red dashed lines give the asymptotes for large and small $d$.}  
\end{figure}

Figure~\ref{fig:HR_PP_plate} shows the normalized (by the HR in isolation) HR of a SiC PP 1 at temperature $ T_1 = 300 \ {\rm K} $ as a function of distance from the mirror.  
For small distance $d$, the HR approaches a finite result, because the mirror does not allow for near field heat transfer modes. Expanding Eq.~\eqref{eq:HR_mirror_plate} for small $d$ yields
\begin{equation}
\lim_{d \ll \lambda_{T_1}} H_{1pp}^{(1pp)} = \frac{2}{3}H_{1pp, vac}^{(1pp)},
\label{eq:HR_mirror_plate_d_small}
\end{equation}
in agreement with the numerical data. This means that $ \frac{1}{3} $ of the HR in vacuum is suppressed by the 
plate. The limit in Eq.~\eqref{eq:HR_mirror_plate_d_small} can be understood by considering a PP as a dipole averaged over three independent orientations with respect to the plate, two parallel and one 
perpendicular. In vacuum, each orientation contributes one third of the total result. In the presence of a mirror plate, the radiation of a dipole parallel to the plate is canceled by its mirror image, 
while the radiation of a perpendicular dipole is doubled~\cite{Barnes1998}. Therefore, we have a doubled contribution of perpendicular orientation which gives two thirds of the radiation of a PP in vacuum. 
With increase of the distance, the HR increases and reaches the global maximum at $ d \approx 3.3\times 10^{-6} \ {\rm m} $ followed by oscillations around the vacuum HR. These may be attributed  to the 
interference effect between the initially radiated waves and the waves reflected from the mirror. The local maxima and minima correspond to the points where emitted and reflected fields 
are in-phase and out-of-phase, respectively. Since the reflected field reaching the particle decreases with separation $ d $, oscillations decrease with $ d $ as well. In the far-field limit 
($ d \gg \lambda_{T_1} $) the HR becomes that in vacuum, because the reflected field is very weak to significantly affect the power dissipated within the particle, i.e., 
\begin{equation}
\lim_{d \gg \lambda_{T_1}} H_{1pp}^{(1pp)} = H_{1pp, vac}^{(1pp)}.
\label{eq:HR_mirror_plate_d_large}
\end{equation}

The partly related  process of emission of excited atoms near interfaces was reviewed in Ref.~\cite{Barnes1998}, and a detailed microscopic description can be found in Ref.~\cite{Yeung1996}. The results of these 
references are in qualitative agreement with our findings.

\section{Heat radiation inside a spherical mirror cavity}
\label{sec:HR_cavity}
\begin{figure}[b]
\includegraphics[width=0.6\linewidth]{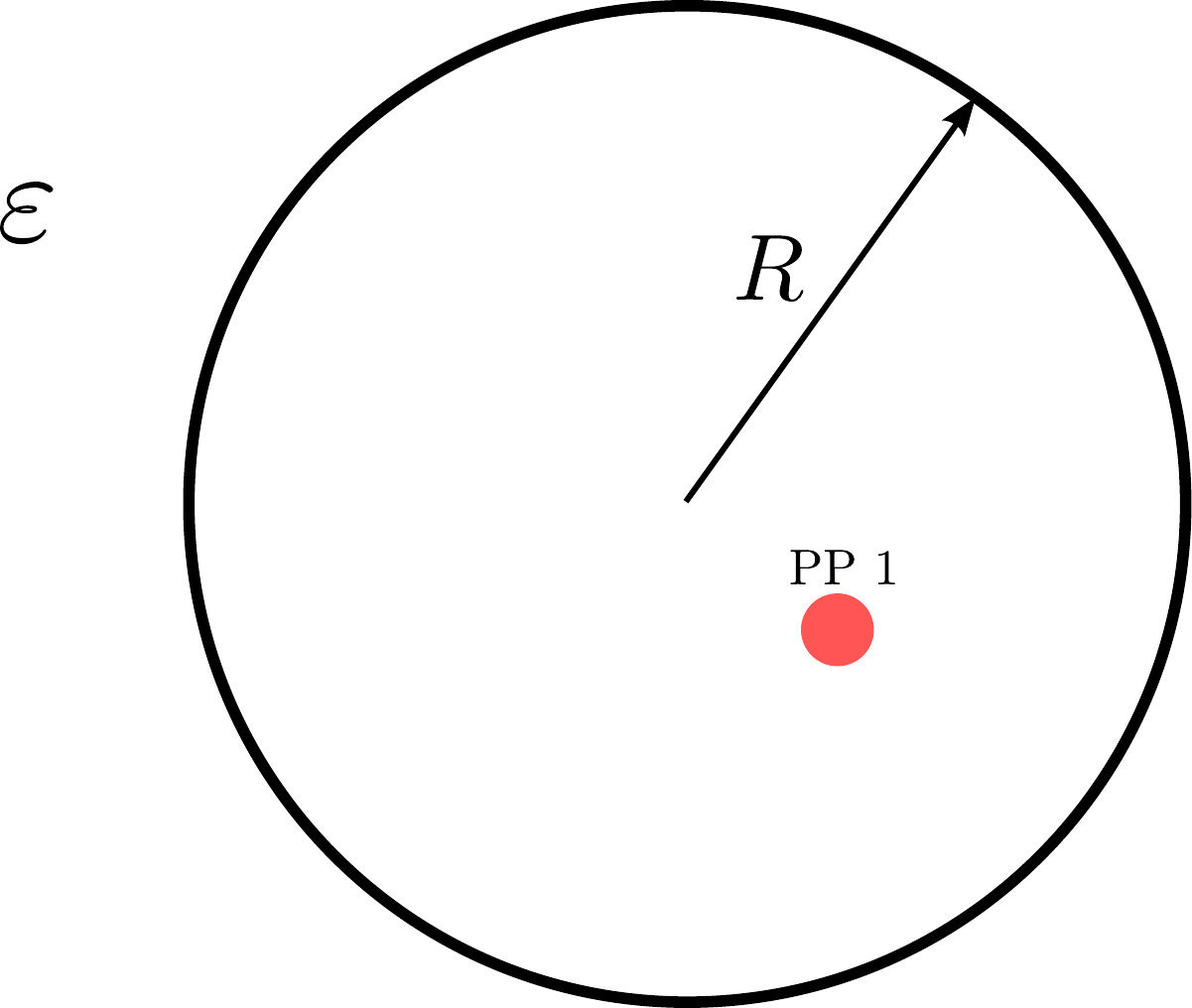}
\caption{\label{fig:config_PP_cavity}A PP inside a spherical cavity. We aim to compute the HR of the particle in the case, where the walls of the cavity perfectly reflect, i.e., $ |\varepsilon| \to \infty $.}  
\end{figure}

Consider a PP placed inside a spherical cavity of radius $ R $ with perfectly reflecting walls as depicted in Fig.~\ref{fig:config_PP_cavity}. The position $\vct{r}_1$ is given in spherical coordinates, with radial, azimuthal, and polar coordinates  $r_1$, $\theta_1$, and $\phi_1$ (i.e., $r_1<R$). It is intuitively clear that the particle, regardless of its material and temperature, does not radiate energy in stationary state. All the waves initially 
emitted by the particle are totally reflected from the walls making the net energy flow zero. To our knowledge, however, there is no mathematical proof of this expectation  in literature, which we provide here. 

The GF of a spherical cavity with both arguments inside the cavity reads (see Appendix~\ref{app:GFs}),
\begin{align}
\notag \mathbb{G}(\vct{r}, \vct{r}') =& \ \mathbb{G}_0(\vct{r}, \vct{r}')\\ 
& + i\sum_{l=1}^{\infty}\sum_{m=-l}^l\sum_{P=M,N}\vct{E}_{Plm}^{\rm reg}(\vct{r}) \otimes \vct{E}_{Pl-m}^{\rm reg}(\vct{r}') \mathcal{T}_l^P, 
\label{eq:GF_cavity_maintext}
\end{align}
where the matrix elements $ \mathcal{T}_l^P $, describing inside-scattering, can be found in Ref.~\cite{Zaheer2010}. Using  symmetry relations, one may show that, at equal arguments,
\begin{align}
\Im \sum_{m=-l}^l \vct{E}_{Plm}^{\rm reg}(\vct{r}_1) \otimes \vct{E}_{Pl-m}^{\rm reg}(\vct{r}_1) = 0.
\end{align}
From it, the imaginary part of the GF, evaluated at equal points inside the cavity, reads as
\begin{align}
&\notag \Im \mathbb{G}(\vct{r}_1, \vct{r}_1) = \ \Im \mathbb{G}_0(\vct{r}_1, \vct{r}_1)\\ 
& + \sum_{P=M,N}\sum_{l=1}^{\infty}\Re\left(\mathcal{T}_l^P\right)\sum_{m=-l}^l\vct{E}_{Plm}^{\rm reg}(\vct{r}_1) \otimes \vct{E}_{Pl-m}^{\rm reg}(\vct{r}_1).
\label{eq:GF_cavity_equal_points_Im}
\end{align}
While this expression is valid for any material of the cavity walls, the perfect mirror limit is obtained by using $ \lim_{|\varepsilon|\to\infty}\Re \mathcal{T}_l^P = -1 $ according to Eq.~\eqref{eq:T_cavity_mirror_real}, and hence
\begin{align}
\notag \Im \mathbb{G}&(\vct{r}_1, \vct{r}_1) = \Im \mathbb{G}_0(\vct{r}_1, \vct{r}_1)\\ 
& - \sum_{l=1}^{\infty}\sum_{m=-l}^l\sum_{P=M,N}\vct{E}_{Plm}^{\rm reg}(\vct{r}_1) \otimes \vct{E}_{Pl-m}^{\rm reg}(\vct{r}_1).
\label{eq:GF_cavity_equal_points_Im_mirror}
\end{align}
Last, recalling that the imaginary part of the free GF can be expanded in \textit{regular} spherical waves as \cite{Kruger2012},
\begin{equation}
\Im \mathbb{G}_0(\vct{r}, \vct{r}') = \sum_{l=1}^{\infty}\sum_{m=-l}^l\sum_{P=M,N}\vct{E}_{Plm}^{\rm reg}(\vct{r}) \otimes \vct{E}_{Pl-m}^{\rm reg}(\vct{r}'),
\label{eq:Im_GF_free_expansion}
\end{equation} 
we note that the two terms in Eq.~\eqref{eq:GF_cavity_equal_points_Im_mirror} cancel, and we 
 finally have for the imaginary part of the GF 
inside a spherical cavity with mirror walls
\begin{equation}
\Im \mathbb{G}(\vct{r}_1, \vct{r}_1) = 0. 
\label{eq:GF_cavity_equal_points_Im_mirror_final}
\end{equation}
This result allows us to make a fundamental statement: the local EM density of states at any point inside a spherical cavity with perfectly conducting walls is zero~\footnote{Initially, the result~\eqref{eq:GF_cavity_equal_points_Im_mirror_final} implies that the \textit{electric} part of the total local EM density of states is zero. It is enough, however, for the \textit{magnetic} part 
of the local EM density of states to be zero as well. Therefore, the total local EM density of states is zero inside a spherical mirror cavity.} . According to 
Eqs.~\eqref{eq:GF_cavity_equal_points_Im_mirror_final} and~\eqref{eq:HR_pp_final}, 
\begin{equation}
H_{1pp}^{(1pp)} = 0,
\label{eq:HR_pp_cavity}
\end{equation}
i.e, a PP placed inside a spherical mirror cavity does not radiate energy, in accordance with conservation of energy.

Formally, we derived the result~\eqref{eq:HR_pp_cavity} \textit{for a PP} (of any material and temperature). It means that the shortest distance from the particle's center to the cavity walls has to be much 
larger than the radius of the particle. Remember that this condition allowed us to neglect multiple reflections from the particle to derive expression~\eqref{eq:HR_pp_final} for the HR. In such a closed 
system, however, the validity for this neglect is under question. Even though the particle is far away from the cavity walls, the reflections from the particle can contribute significantly, because the 
initially radiated field always comes back to it.

\section{Conclusion}
\label{sec:Conclusion}
In this paper, we studied heat radiation of a point particle and heat transfer between two point particles in the presence of an arbitrary system of objects. We applied the PP limit for the radiating and 
absorbing particles to derive general formulas for HR and HT. These formulas, whose main element is the Green's function of the objects surrounding the PPs, are much simpler 
than the general expressions for HR and HT. They open up a powerful route for (numerical) studies of HR and HT for PPs in various systems, where the surrounding objects  can 
be of arbitrary shape, size, and material.

The HT can be dramatically enhanced by a sphere placed between the two PPs, where the strongest effect was seen if all materials are SiC. For a well-conducting sphere, we also found a significant enhancement, which is a pure diffraction effect: the view factor for the given configuration is exactly zero. The self-emission of a PP in front of a semi-infinite mirror plate is finite at all distances. It approaches $ \frac{2}{3} $ of the vacuum result for small $d$, and shows oscillations in the region  where the separation distance is comparable to the thermal wavelength. The HR of a PP placed inside a spherical cavity with perfectly conducting walls is exactly zero.

While there are already a lot of studies of HR and HT in many-body systems (where all  objects are small particles), the research in the field of many-body HR and HT still provides many open and interesting questions. The formalism presented in this paper could be used to study the HT between two PPs in the presence of a multilayered sphere. The problem of the self-emission of a PP in front of a mirror plate can be straightforwardly extended to the HR of a PP or 
HT between two PPs in the presence of a plate made of various materials, or inside a cavity made by two parallel plates. Future work may also study the  HT between two PPs in the presence of a cylinder, where interesting phenomena due to surface waves may be expected.

\begin{acknowledgements}
We thank G. Bimonte, T. Emig, N. Graham,
R. L. Jaffe and M. Kardar for useful discussions. 
This work was supported by MIT-Germany
Seed Fund Grant No. 2746830 and 
Deutsche Forschungsgemeinschaft (DFG)
Grants No. KR 3844/2-1 and KR 3844/3-1.
K.A. also acknowledges the support by International Max Planck Research School (IMPRS) for Condensed Matter Science in Stuttgart.
\end{acknowledgements}

\begin{appendix}

\section{Electromagnetic operators}
\label{app:EM_operators}
For an abstract frequency-dependent operator $ \mathbb{A}(\omega) $, its position space representation is defined as
\begin{equation}
\mathbb{A}(\omega, \vct{r}, \vct{r'}) \equiv A_{ij}(\omega, \vct{r}, \vct{r'}) = \bra{\vct{r}}\mathbb{A}(\omega)\ket{\vct{r'}},
\label{eq:operator_ps_def}
\end{equation}
where $ \ket{\vct{r}} $ is an eigenstate of the position operator. In general, $ \mathbb{A}(\omega, \vct{r}, \vct{r'}) $ is a $ 3 \times 3 $ position-dependent matrix 
containing arbitrary operations, e.g., derivatives. The dependence on $ \omega $ is not made explicitly in frequency-dependent operators. In this paper, 
operators are assumed to be initially in position space representation. Operator products involve matrix multiplication, as well as integration over the common spatial 
argument:
\begin{equation}
\mathbb{A}\mathbb{B} \equiv (AB)_{ik}(\vct{r}, \vct{r'}) = \int d^3r'' A_{ij}(\vct{r}, \vct{r''})B_{jk}(\vct{r''}, \vct{r'}),
\label{eq:operator_product_def}
\end{equation}
where the summation over repeated indices is understood. The trace of an operator is both over the matrix indices and the spatial arguments. 

Important operators for our considerations are potential $ \mathbb{V} $ introduced by the objects, Green's function (GF) $ \mathbb{G} $ of the system and the scattering operator 
$ \mathbb{T} $. The potential is defined as
\begin{equation}
\mathbb{V} = \frac{\omega^2}{c^2}(\bbeps-\mathbb{I})+\boldsymbol{\nabla}\times\left(\mathbb{I}-\frac{1}{\bbmu}\right)\boldsymbol{\nabla}\times
\label{eq:potential_def}
\end{equation}
and GF obeys the Helmholtz equation
\begin{equation}
\left[ \mathbb{H}_0-\mathbb{V}-\frac{\omega^2}{c^2}\mathbb{I}\right]\mathbb{G}(\vct{r},\vct{r}')=\mathcal{I}\delta^{(3)}(\vct{r}-\vct{r}'),
\label{eq:Helmholtz_eq}
\end{equation}
where $\mathbb{H}_0=\boldsymbol{\nabla}\times\boldsymbol{\nabla}\times$ and $ \mathcal{I} $ is the $ 3\times3 $ identity matrix~\footnote{We use symbol $ \mathcal{I} $ for the $ 3\times3 $ identity 
matrix and $ \mathbb{I} = \mathcal{I}\delta^{(3)}(\vct{r}-\vct{r}') $ for the identity operator.}. The free GF (see Eq.~\eqref{eq:GF_free}) is the 
solution of Eq.~\eqref{eq:Helmholtz_eq} with $ \mathbb{V} = 0 $. The scattering operator can be defined in the context of Lippmann-Schwinger equation~\cite{Lippmann1950} and reads:
\begin{equation}
\mathbb{T}=\mathbb{V}\frac{1}{\mathbb{I}-\mathbb{G}_0\mathbb{V}}.
\label{eq:T_def}
\end{equation}
Using Eqs.~\eqref{eq:Helmholtz_eq} and~\eqref{eq:T_def}, one can find the following relation between $ \mathbb{G} $ and $ \mathbb{T} $:
\begin{equation}
\mathbb{G} = \mathbb{G}_0+\mathbb{G}_0\mathbb{T}\mathbb{G}_0.
\label{eq:GF_via_T}
\end{equation}

\section{Green's functions}
\label{app:GFs}
GF of an object is usually found using Eq.~\eqref{eq:GF_via_T}. One typically applies expansion of the free GF in partial waves of an appropriate basis to have the result in terms of the waves and the scattering 
matrix of an object~\cite{Golyk2012, Kruger2012, Tsang2000, Rahi2009}.

\subsection{Free space}
The GF of free space can be written in closed form~\cite{Tsang2000, VanBladel1961, Yaghjian1980, Weiglhofer1989}:
\begin{align}
\notag \mathbb{G}_0(\vct{r}, \vct{r}') =& \ -\frac{1}{3k^2}\mathcal{I}\delta^{(3)}(\vct{r}-\vct{r}')\\
\notag &  + \frac{e^{ikd}}{4\pi k^2d^5}\Big[d^2(-1+ikd+k^2d^2)\mathcal{I}\\
& + (3-3ikd-k^2d^2)(\vct{r}-\vct{r}')\otimes(\vct{r}-\vct{r}')\Big],
\label{eq:GF_free}
\end{align}
where $ d = |\vct{r}-\vct{r}'| $ is the distance between the points and the symbol $ \otimes $ denotes the dyadic product. Note the delta function term, which contributes to the field at the source 
region ~\cite{Tsang2000, VanBladel1961, Yaghjian1980, Weiglhofer1989, Nevel2004}. As expected from translational invariance of the observation points, the free GF is a function of $ \vct{r}-\vct{r}' $.

\subsection{Sphere}
Firstly, we introduce spherical waves and scattering matrix of a sphere according to Ref.~\cite{Kruger2012}. We consider that the center of the sphere is located at the origin of the coordinate system. The 
spherical waves are solutions of the wave equation in spherical coordinates 
($ r $, $ \theta $, $ \phi $):
\begin{align}
& \vct{E}^{\rm reg}_{Mlm} = \sqrt{(-1)^m k}\frac{1}{\sqrt{l(l+1)}}j_l\left(kr\right)\nabla\times\vct{r}Y_l^m(\theta,\phi),\label{eq:spherical_waves_Mreg}\\
& \vct{E}^{\rm out}_{Mlm} = \sqrt{(-1)^m k}\frac{1}{\sqrt{l(l+1)}}h_l\left(kr\right)\nabla\times\vct{r}Y_l^m(\theta,\phi),\label{eq:spherical_waves_Mout}\\
& \vct{E}^{\rm reg}_{Nlm} = \frac{1}{k}\nabla\times\vct{E}^{\rm reg}_{Mlm},\label{eq:spherical_waves_Nreg}\\
& \vct{E}^{\rm out}_{Nlm} = \frac{1}{k}\nabla\times\vct{E}^{\rm out}_{Mlm}.\label{eq:spherical_waves_Nout}
\end{align}
Here, $ l = 1,\ 2,\ \dots $; $ m = -l,\ -(l-1),\dots,\ 0,\dots,\  (l-1),\ l $; indices $ M $ and $ N $ denote magnetic and electric polarizations, respectively; the superscript \enquote{$ {\rm reg} $} means that the wave 
is regular at the origin, while the superscript \enquote{$ {\rm out} $} means that the wave is singular at the origin, i.e., it is the outgoing wave; 
\begin{equation}
Y_l^m(\theta, \phi) = \sqrt{\frac{(2l+1)}{4\pi}\frac{(l-m)!}{(l+m)!}}P_l^m(\cos{\theta})e^{im\phi} 
\label{eq:spherical_harmonics}
\end{equation}
are spherical harmonics, where $ P_l^m(\cos{\theta}) $ is the Legendre function; $ j_l(kr) $ is the spherical Bessel function of order $ l $ and $ h_l(kr) $ is the spherical Hankel function of the first kind
of order $ l $. The scattering matrix of a sphere can be defined in the context of Lippmann-Schwinger equation~\cite{Kruger2012, Rahi2009}. For the case where the matrix relates the incident and the scattered 
fields \textit{outside a sphere}, it reads as
\begin{equation}
\mathcal{T}_{\mu \mu'} = i \int d^3r \int d^3r' \vct{E}_{\sigma(\mu)}^{{\rm reg}}(\vct{r})\mathbb{T}(\vct{r},\vct{r}')\vct{E}_{\mu'}^{\rm reg}(\vct{r}'),
\label{eq:scattering_matrix_sphere_def}
\end{equation}
where $ \mathbb{T}(\vct{r},\vct{r}') $ is the scattering operator of a sphere, $ \mu = \{P, l, m\} $ ($ P $ denotes polarization, magnetic $ M $ or electric $ N $) and $ \sigma(\mu) = \{P, l, -m\} $. 

The matrix elements can be obtained by solving the boundary conditions problem~\cite{Bohren2004, Tsang2000}. For a sphere with isotropic and local $ \varepsilon $ and $ \mu $, the matrix is diagonal and 
independent on $ m $, $ \mathcal{T}_{\mu \mu'} \equiv \mathcal{T}_{ll'mm'}^{PP'} = \mathcal{T}_{l}^{P}\delta_{PP'}\delta_{ll'}\delta_{mm'} $. The matrix elements $ \mathcal{T}_{l}^{P} $ for a homogeneous 
sphere of radius $ R $ (see Fig.~\ref{fig:config_PPs_sphere}) are given by~\cite{Bohren2004, Kruger2012, Tsang2000}
\begin{align}
& \mathcal{T}_{l}^M=-\frac{\mu j_l(\tilde R^*)\frac{d}{d R^*}[R^*j_l(R^*)]-j_l(R^*)\frac{d}{d \tilde R^*}[\tilde R^*j_l(\tilde R^*)]}{\mu j_l(\tilde R^*)\frac{d}{d R^*}[R^*h_l(R^*)]-h_l(R^*)\frac{d}{d \tilde R^*}[\tilde R^*j_l(\tilde R^*)]},\label{eq:TM_sphere}\\
& \mathcal{T}_{l}^N=-\frac{\varepsilon j_l(\tilde R^*)\frac{d}{d R^*}[R^*j_l(R^*)]-j_l(R^*)\frac{d}{d \tilde R^*}[\tilde R^*j_l(\tilde R^*)]}{\varepsilon j_l(\tilde R^*)\frac{d}{d R^*}[R^*h_l(R^*)]-h_l(R^*)\frac{d}{d \tilde R^*}[\tilde R^*j_l(\tilde R^*)]},\label{eq:TN_sphere}
\end{align}
where $ R^* = kR $ and $ \tilde R^* = \sqrt{\varepsilon \mu}kR $. In the limit of perfect conductivity (or 
reflectivity), the scattering matrix simplifies to
\begin{align}
& \lim_{|\varepsilon| \to \infty}\mathcal{T}_{l}^M= - \frac{j_l(R^*)}{h_l(R^*)},\label{eq:TM_sphere_mirror}\\
& \lim_{|\varepsilon| \to \infty}\mathcal{T}_{l}^N= - \frac{\frac{d}{dR^*}\left[R^*j_l(R^*)\right]}{\frac{d}{dR^*}\left[R^*h_l(R^*)\right]}.\label{eq:TN_sphere_mirror}
\end{align}

The GF of a sphere can be evaluated using Eq.~\eqref{eq:GF_via_T} and expansion of the free GF in spherical waves~\cite{Kruger2012, Tsang2000, Rahi2009}. For its both position arguments lying outside the sphere, 
the GF reads as
\begin{align}
\notag \mathbb{G}(\vct{r}, \vct{r}') =& \ \mathbb{G}_0(\vct{r}, \vct{r}')\\ 
& + i\sum_{l=1}^{\infty}\sum_{m=-l}^l\sum_{P=M,N}\vct{E}_{Plm}^{\rm out}(\vct{r}) \otimes \vct{E}_{Pl-m}^{\rm out}(\vct{r}') \mathcal{T}_l^P, 
\label{eq:GF_sphere}
\end{align}
where matrix elements $ \mathcal{T}_l^P $ are given by Eqs.~\eqref{eq:TM_sphere} and~\eqref{eq:TN_sphere}. For the free GF in formula~\eqref{eq:GF_sphere}, it is reasonable to use closed form 
expression~\eqref{eq:GF_free} in spherical coordinates. We note again that GF~\eqref{eq:GF_sphere}, as well as matrix elements~\eqref{eq:scattering_matrix_sphere_def}, are valid for only outside-outside 
scattering, such that both observation points lie outside the sphere.

\subsection{Small sphere and point particle}
The GF of a small homogeneous isotropic nonmagnetic sphere can be readily obtained using Eq.~\eqref{eq:GF_via_T} and the scattering operator~\eqref{eq:T_ss}. We have
\begin{align}
\notag \mathbb{G}(\vct{r}, \vct{r}') =& \ \mathbb{G}_0(\vct{r}, \vct{r}')\\
& + 3k^2\frac{\varepsilon-1}{\varepsilon+2}\int_{V_{ss}} d^3r''\mathbb{G}_0(\vct{r}, \vct{r}'')\mathbb{G}_0(\vct{r}'', \vct{r}'),
\label{eq:GF_ss}
\end{align}
where the integration runs over the volume $ V_{ss} $ of the sphere. This GF has no restriction on its arguments.

If a small sphere reduces to a PP, such that the observation points are far away from the sphere, formula~\eqref{eq:GF_ss} can be further simplified. The integration variable inside free 
GFs in the integral of expression~\eqref{eq:GF_ss} can be replaced by the fixed coordinate $ \vct{r}_0 $ corresponding to the center of the PP with radius $ R $, and we obtain
\begin{equation}
\mathbb{G}(\vct{r}, \vct{r}') = \mathbb{G}_0(\vct{r}, \vct{r}') + 4\pi k^2 \alpha \mathbb{G}_0(\vct{r}, \vct{r}_0)\mathbb{G}_0(\vct{r}_0, \vct{r}'),
\label{eq:GF_pp}
\end{equation}
where $ \alpha $ is electrical dipole polarizability of the particle given in Eq.~\eqref{eq:polarazability}. An expression for the GF of a point dipole similar to Eq.~\eqref{eq:GF_pp} can be found in 
Ref.~\cite{Carminati2006}.

\subsection{Plate}
We consider a semi-infinite homogeneous isotropic plate occupying the space $ z \leqslant 0 $ (see Fig.~\ref{fig:config_PP_plate}) and aim to find the GF with both arguments lying outside the plate. Firstly, we introduce 
the plane waves similar to those defined in Refs.~\cite{Kruger2012, Tsang2000, Rahi2009}:
\begin{align}
& \vct{M}_{\vct{k}_\perp}(\vct{x}_\perp, z) = \frac{1}{|\vct{k}_\perp|}(\hat{\vct{x}}k_y-\hat{\vct{y}}k_x)e^{i\vct{k}\cdot \vct{r}}, \label{eq:plane_waves_M}\\
& \vct{N}_{\vct{k}_\perp}^{\pm}(\vct{x}_\perp, z) = \frac{1}{k|\vct{k}_\perp|}(\pm \hat{\vct{x}}k_xk_z \pm \hat{\vct{y}}k_yk_z + \hat{\vct{z}}k_\perp^2)e^{i\vct{k}\cdot \vct{r}},\label{eq:plane_waves_N}
\end{align}
where $ \vct{r} $ is the radius vector, $ \vct{k} = (\vct{k}_\perp, k_z)^T $ and $ k_z = \sqrt{k^2-\vct{k}_\perp^2} $. The scattering matrix of a plate can be written in terms of Fresnel reflection 
coefficients~\cite{Kruger2012, Muller2017, Jackson1999}
\begin{align}
& r^M = \frac{\mu\sqrt{\frac{\omega^2}{c^2}-k_\perp^2}-\sqrt{\varepsilon\mu\frac{\omega^2}{c^2}-k_\perp^2}}{\mu\sqrt{\frac{\omega^2}{c^2}-k_\perp^2}+\sqrt{\varepsilon\mu\frac{\omega^2}{c^2}-k_\perp^2}},\label{eq:Fresnel_coeff_M}\\
& r^N = \frac{\varepsilon\sqrt{\frac{\omega^2}{c^2}-k_\perp^2}-\sqrt{\varepsilon\mu\frac{\omega^2}{c^2}-k_\perp^2}}{\varepsilon\sqrt{\frac{\omega^2}{c^2}-k_\perp^2}+\sqrt{\varepsilon\mu\frac{\omega^2}{c^2}-k_\perp^2}},\label{eq:Fresnel_coeff_N}
\end{align}
which in the case of a perfect mirror plate simplify to
\begin{align}
& \lim_{|\varepsilon| \to \infty}r^M = -1,\label{eq:Fresnel_coeff_M_mirror}\\
& \lim_{|\varepsilon| \to \infty}r^N = 1.\label{eq:Fresnel_coeff_N_mirror}
\end{align}
Similar to the case of a sphere, the GF of a plate can be obtained using the expansion of the free GF in the plane waves~\cite{Kruger2012, Tsang2000, Rahi2009} and formula~\eqref{eq:GF_via_T}. We have
\begin{align}
\notag \mathbb{G}(\vct{r}, \vct{r}') =& \ \mathbb{G}_0(\vct{r}, \vct{r}') + \frac{i}{8\pi^2}\int d^2k_\perp\frac{1}{k_z}\\ 
\notag & \times \Big[\vct{M}_{\vct{k}_\perp}(\vct{x}_\perp, z)\otimes \vct{M}_{\vct{k}_\perp}(-\vct{x}'_\perp, z')r^M\\ 
& + \vct{N}_{\vct{k}_\perp}^-(\vct{x}_\perp, z)\otimes \vct{N}_{\vct{k}_\perp}^+(-\vct{x}'_\perp, z')r^N\Big].
\label{eq:GF_plate}
\end{align}

\subsection{Spherical cavity}
We consider an object occupying the whole space except centered at the origin spherical cavity of radius $ R $ (see Fig.~\ref{fig:config_PP_cavity}). We aim to find the GF inside the cavity. Due to spherical 
symmetry of the system, one uses the spherical waves introduced above to find the GF. In contrast to the case of a sphere, the scattering matrix of the cavity walls with scattering operator 
$ \mathbb{T}(\vct{r}, \vct{r}') $ contains outgoing waves
\begin{equation}
\mathcal{T}_{\mu \mu'} = i \int d^3r \int d^3r' \vct{E}_{\sigma(\mu)}^{{\rm out}}(\vct{r})\mathbb{T}(\vct{r},\vct{r}')\vct{E}_{\mu'}^{\rm out}(\vct{r}'),
\label{eq:scattering_matrix_cavity_def}
\end{equation}
and one has to use different pieces of expansion of the free GF~\cite{Kruger2012, Tsang2000, Rahi2009} in formula~\eqref{eq:GF_via_T}. We consider that $ \varepsilon $ and $ \mu $ of the cavity walls are 
isotropic and homogeneous, such that $ \mathcal{T}_{\mu \mu'} \equiv \mathcal{T}_{ll'mm'}^{PP'} = \mathcal{T}_{l}^{P}\delta_{PP'}\delta_{ll'}\delta_{mm'} $~\cite{Zaheer2010}. The GF reads as
\begin{align}
\notag \mathbb{G}(\vct{r}, \vct{r}') =& \ \mathbb{G}_0(\vct{r}, \vct{r}')\\ 
& + i\sum_{l=1}^{\infty}\sum_{m=-l}^l\sum_{P=M,N}\vct{E}_{Plm}^{\rm reg}(\vct{r}) \otimes \vct{E}_{Pl-m}^{\rm reg}(\vct{r}') \mathcal{T}_l^P. 
\label{eq:GF_cavity}
\end{align}

The matrix elements $ \mathcal{T}_l^P $ can be found in Ref.~\cite{Zaheer2010}. In the case of perfectly conducting walls, they are inverse of those for perfectly conducting sphere:
\begin{align}
& \lim_{|\varepsilon| \to \infty}\mathcal{T}_{l}^M= - \frac{h_l(R^*)}{j_l(R^*)},\label{eq:TM_cavity_mirror}\\
& \lim_{|\varepsilon| \to \infty}\mathcal{T}_{l}^N= - \frac{\frac{d}{dR^*}\left[R^*h_l(R^*)\right]}{\frac{d}{dR^*}\left[R^*j_l(R^*)\right]}.\label{eq:TN_cavity_mirror}
\end{align}
Since $ R^* = kR $ is real, $ j_l(R^*) $ is real and $ \Re h_l(R^*) = j_l(R^*) $. Therefore, the above matrix elements have the following remarkable property:
\begin{equation}
\Re \lim_{|\varepsilon| \to \infty}\mathcal{T}_{l}^M = \Re \lim_{|\varepsilon| \to \infty}\mathcal{T}_{l}^N = -1.
\label{eq:T_cavity_mirror_real}
\end{equation}

\section{Correspondence between the scattering operator of a small sphere and the scattering matrix}
\label{app:ss_operator_matrix_correspondence}
In Sec.~\ref{sec:HR_HT_PP_limit}, we derived the scattering operator of a small sphere. In this appendix, we check its expression. In contrast to the scattering~\textit{operator}, the 
scattering~\textit{matrix} of a sphere is well known [see Eqs.~\eqref{eq:TM_sphere} and~\eqref{eq:TN_sphere}], and one can hence obtain its form in the small sphere limit. Using the definition of the scattering 
matrix in Eq.~\eqref{eq:scattering_matrix_sphere_def}, we show that the scattering operator for a small sphere given by Eq.~\eqref{eq:T_ss} indeed gives the corresponding scattering matrix.

In Ref.~\cite{Kruger2012}, it was found that in the small sphere limit only the $\mathcal{T}_1^N$ element is relevant and it reads as
\begin{equation}
\mathcal{T}_1^N = i\frac{2(\varepsilon-1)}{3(\varepsilon+2)}k^3R^3,
\label{eq:scattering_matrix_ss}
\end{equation}
where $ R $ is the sphere's radius.
On the other hand, the scattering operator~\eqref{eq:T_ss} gives the following scattering matrix according to Eq.~\eqref{eq:scattering_matrix_sphere_def}
\begin{equation}
\mathcal{\widetilde{T}}_{\mu \mu'} = 3ik^2\frac{\varepsilon-1}{\varepsilon+2} \int_{V_{ss}} d^3r \vct{E}_{\sigma(\mu)}^{{\rm reg}}(\vct{r})\cdot\vct{E}_{\mu'}^{\rm reg}(\vct{r}),
\label{eq:scattering_matrix_from_T_ss}
\end{equation}
where the integration runs over the volume $ V_{ss} $ of the sphere. We have to show that all the elements $ \mathcal{\widetilde{T}}_{\mu \mu'} $ except $ \mathcal{\widetilde{T}}_1^N $ are zeros and that 
$ \mathcal{\widetilde{T}}_1^N = \mathcal{T}_1^N $ given in Eq.~\eqref{eq:scattering_matrix_ss}. Performing the curls in Eqs.~\eqref{eq:spherical_waves_Mreg} and~\eqref{eq:spherical_waves_Nreg}, we 
find~\cite{Tsang2000}
\begin{align}
\notag &\vct{E}_{Mlm}^{\rm reg}(kr, \theta, \phi) = \frac{\sqrt{(-1)^m k}}{\sqrt{l(l+1)}}j_l(kr)\Big[\hat{\theta}\frac{im}{\sin{\theta}}Y_l^m(\theta, \phi)\\
& -\hat{\phi}\frac{d}{d\theta}Y_l^m(\theta, \phi)\Big],\label{eq:spherical_waves_Mreg_curled}\\
\notag &\vct{E}_{Nlm}^{\rm reg}(kr, \theta, \phi) = \frac{\sqrt{(-1)^m k}}{\sqrt{l(l+1)}}\Big\{\hat{r}\frac{l(l+1)j_l(kr)}{kr}Y_l^m(\theta, \phi)\\
& +\frac{1}{kr}\frac{d}{d(kr)}[krj_l(kr)]\Big[\hat{\theta}\frac{d}{d\theta}Y_l^m(\theta, \phi)+\hat{\phi}\frac{im}{\sin{\theta}}Y_l^m(\theta, \phi)\Big]\Big\}.\label{eq:spherical_waves_Nreg_curled}
\end{align}
We consider $ k \lessapprox k_T = \frac{\omega_T}{c} $, because larger wave vectors are not relevant for HR or HT. Since the spatial arguments of the waves in Eq.~\eqref{eq:scattering_matrix_from_T_ss} lie within the 
volume of a \textit{small} sphere, we have $ kr \ll 1 $, which allows us to use the small argument limit of spherical Bessel functions~\cite{Tsang2000}:
\begin{equation}
\lim_{kr \ll 1} j_l(kr) \approx \frac{1}{1\cdot3\cdot5\dots(2l+1)}(kr)^l.
\label{eq:sph_Bessel_small_arg_limit}
\end{equation}
Applying this limit in Eqs.~\eqref{eq:spherical_waves_Mreg_curled} and~\eqref{eq:spherical_waves_Nreg_curled}, it is easy to see that only the waves $ \vct{E}_{N1m}^{\rm reg} $ are relevant. Therefore, all 
the matrix elements (except $ \mathcal{\widetilde{T}}_1^N $) are negligible compared to $ \mathcal{\widetilde{T}}_1^N $ and hence can be considered zero. It remains to show that 
$ \mathcal{\widetilde{T}}_1^N $ = $ \mathcal{T}_1^N $. We write $ \vct{E}_{N1m}^{\rm reg} $ explicitly in the limit $ kr \ll 1 $:
\begin{align}
\notag &\vct{E}_{N1m}^{\rm reg}(kr, \theta, \phi) \approx \sqrt{(-1)^m k}\frac{\sqrt{2}}{3}\Big\{\hat{r}Y_1^m(\theta, \phi)\\
& +\hat{\theta}\frac{d}{d\theta}Y_1^m(\theta, \phi) + \hat{\phi}\frac{im}{\sin{\theta}}Y_1^m(\theta, \phi)\Big\}.
\label{eq:EN1m_small_argument_limit}
\end{align}
Substituting expression~\eqref{eq:EN1m_small_argument_limit} into Eq.~\eqref{eq:scattering_matrix_from_T_ss}, it is straightforward to show that $ \mathcal{\widetilde{T}}_1^N $ = $ \mathcal{T}_1^N $. Therefore, 
the scattering operator of a small sphere~\eqref{eq:T_ss} gives the correct scattering matrix~\eqref{eq:scattering_matrix_ss}.

\end{appendix}

%

\end{document}